\documentclass[usenatbib]{mn2e}
\usepackage{epsfig}

\title[Star formation rate and metallicity of DLAs in cosmological simulations]
{Star formation rate and metallicity of damped Lyman-$\alpha$
absorbers in cosmological SPH simulations}

\author[K. Nagamine, V. Springel, and L. Hernquist]
  {K.~Nagamine,$^1$\thanks{Email: knagamin@cfa.harvard.edu}
  V.~Springel,~$^2$\thanks{Email: volker@mpa-garching.mpg.de} and
  L.~Hernquist$^1$\thanks{Email: lars@cfa.harvard.edu}
   \vspace{0.3cm}\\ 
  $^1$Harvard-Smithsonian Center for Astrophysics,
  60 Garden Street, Cambridge, MA 02138, U.S.A. \\
  $^2$Max-Planck-Institut f\"{u}r Astrophysik,
  Karl-Schwarzschild-Stra\ss{}e 1, 85740 Garching bei M\"{u}nchen,
  Germany}

\pagerange{\pageref{firstpage}--\pageref{lastpage}}
\pubyear{2003}

\newcommand{\avg}[1]{\left\langle#1\right\rangle}
\newcommand{\rp}[1]{\left(#1\right)}
\newcommand{\Lam}{\Lambda}

\newcommand{\kpc}{{\rm kpc}}
\newcommand{\pc}{{\rm pc}}
\newcommand{\cm}{\rm cm}
\newcommand{\yr}{\rm yr}
\newcommand{\Gyr}{\rm Gyr}
\newcommand{\kms}{\rm km\,s^{-1}}
\newcommand{\K}{\rm K}
\newcommand{\ergs}{\rm ergs}
\newcommand{\Msun}{\rm M_{\odot}}

\newcommand{\Zsun}{Z_{\odot}}
\newcommand{\hinv}{h^{-1}}
\newcommand{\himpc}{\hinv{\rm\,Mpc}}
\newcommand{\hikpc}{\hinv{\rm\,kpc}}
\newcommand{\himsun}{\hinv{\Msun}}

\newcommand{\Om}{\Omega_{\rm m}}
\newcommand{\Ol}{\Omega_{\Lam}}
\newcommand{\Ob}{\Omega_{\rm b}}
\newcommand{\OHI}{\Omega_{\rm HI}}
\newcommand{\HI}{H{\sc i}\,\,}
\newcommand{\NHI}{{\rm N_{{\rm HI}}}}

\newcommand{\XH}{X_{\rm H}}

\newcommand{\Mhalo}{M_{\rm halo}}
\newcommand{\Lbox}{{\rm L_{box}}}
\newcommand{\Sigstar}{\Sigma_{\rm STAR}}
\newcommand{\Sigsfr}{\Sigma_{\rm SFR}}
\newcommand{\Sigism}{\Sigma_{\rm ISM}}
\newcommand{\CII}{C{\sc ii}}
\newcommand{\tstar}{t_{\star}}
\newcommand{\rhostar}{\rho_{\star}}
\newcommand{\rhoth}{\rho_{\rm th}}
\newcommand{\dd}{{\rm d}}
\newcommand{\tzstar}{t_0^{\star}}
\newcommand{\esn}{\epsilon_{\rm SN}}
\newcommand{\Mwdot}{\dot{M}_W}
\newcommand{\Msdot}{\dot{M}_{\star}}
\newcommand{\usn}{u_{\rm SN}}
\newcommand{\Nh}{N_{\rm h}}
\newcommand{\mpro}{m_{\rm p}}

\newcommand{\beq}{\begin{eqnarray}}
\newcommand{\eeq}{\end{eqnarray}}

\begin{document}

\maketitle

\label{firstpage}


\begin{abstract}
We study the distribution of the star formation rate (SFR) and metallicity
of damped Lyman-$\alpha$ absorbers (DLAs) in the redshift range $z=0-4.5$
using cosmological smoothed particle hydrodynamics (SPH) simulations 
of the $\Lam$ cold dark matter model.  Our simulations include
standard radiative cooling and heating with a uniform UV background,
star formation, supernova (SN) feedback, as well as a phenomenological
model for feedback by galactic winds. The latter allows us to examine,
in particular, the effect of galactic outflows on the distribution of
the SFR and metallicity of DLAs.  We employ a ``conservative entropy''
formulation of SPH which alleviates numerical overcooling effects that
affected earlier simulations. In addition, we utilise a series of
simulations of varying boxsize and particle number to investigate the
impact of numerical resolution on our results.  

We find that there is a positive correlation between the projected
stellar mass density and the neutral hydrogen column density ($\NHI$)
of DLAs for high $\NHI$ systems, and that there is a good
correspondence in the spatial distribution of stars and DLAs in the
simulations.  The evolution of typical star-to-gas mass ratios in DLAs
can be characterised by an increase from about 2 at $z=4.5$ to 3 at
$z=3$, to 10 at $z=1$, and finally to 20 at $z=0$.  We also find that
the projected SFR density in DLAs follows the Kennicutt law well at
all redshifts, and the simulated values are consistent with the recent
observational estimates of this quantity by \citet{Wol03a, Wol03b}.
The rate of evolution in the mean metallicity of simulated DLAs as a
function of redshift is mild, and is consistent with the rate
estimated from observations.  The predicted metallicity of DLAs is
generally sub-solar in our simulations, and there is a significant
scatter in the distribution of DLA metallicity for a given
$\NHI$. However, we find that the median metallicity of simulated DLAs
is close to that of Lyman-break galaxies, and is higher than the
values typically observed for DLAs by nearly an order of magnitude.
This discrepancy with observations could be due to an inadequate
treatment of the SN feedback or the multiphase structure of the gas
in our current simulations. Alternatively, the current observations 
might be missing the majority of the high metallicity DLAs due to 
selection effects.
\end{abstract}



\section{Introduction}
\label{section:intro}

Damped Lyman-$\alpha$ absorbers (defined as quasar absorption systems
with column density $\NHI>2\times 10^{20} \cm^{-2}$) have similar
neutral hydrogen column density as galactic disks in the Local
Universe. They are high concentrations of neutral gas which could form
stars and eventually evolve into galaxies that we see today
\citep{Wol86}.  While the most numerous absorbers are those with low
column densities, the high column densities of DLAs more than
compensate for their relative paucity when the relative contribution
to the total neutral hydrogen (\HI) mass density is considered. It is
thought that DLAs dominate the \HI gas content of the Universe at
$z\sim 3$ \citep{Lan93, Wol95, Sto00}, and thus they are serving as an
important reservoir of neutral gas for star formation at high
redshift.  Therefore, studying the physical properties of DLAs, such
as the distribution of their star formation rates and metallicity, will
provide us with important complementary information to those provided
by the emitted light from stars in high-redshift galaxies.

It has become clear from recent observational \citep[e.g.][]{Ade98,
Ste99} and theoretical \citep[e.g.][]{Mo96, Bau98, Mo99, KHW99, Kau99,
Nag02, Wei02} studies of Lyman-break galaxies (LBGs) at $z\sim 3-4$,
that the assembly of galaxies is actively going on at $z\sim 3$,
consistent with what is expected from the hierarchical structure
formation paradigm based on a cold dark matter (CDM) model. The star
formation rates of LBGs at $z=3-4$ are now commonly observed using
emission lines such as H$\alpha$, or by fitting the entire spectrum with
population synthesis models \citep[e.g.][]{Sha01}.  These
measurements of SFRs typically give values of $\sim 50 ~\Msun\yr^{-1}$
for LBGs, allowing one to place a lower limit on the cosmic star
formation rate by combining the information on the SFR and number density
of LBGs.

Furthermore, many theoretical and observational studies suggest that
the cosmic star formation rate rises towards high redshift, even
beyond $z=3$ \citep[e.g.][]{Pas98, Bla99, Nag01a, Lan02, SH03b,
Her03}.  Therefore, the conversion of gas into stars is taking place
at a significant rate in these high-redshift galaxies at $z\geq 3$.
Studying this conversion process can help us understand the galaxy
formation better, and possibly constrain structure formation scenarios
such as the hierarchical CDM models.

However, estimating the star formation rate in DLAs has proven to be
very difficult, because the stellar counterparts of DLAs cannot be
detected in emission due to their faintness.  \citet{Wol03a, Wol03b}
have therefore adopted a different approach and recently estimated the
SFR in DLAs in the redshift range $z=3-4$ by utilising the \CII$^*$
1335.7 absorption line. They infer the heating rate by equating it to
the cooling rate measured from \CII$^*$ absorption. Since the heating
rate is proportional to the SFR, they can then deduce the SFR from the
estimated heating rate. The values of SFR they measure by this method
roughly agree with the Kennicutt law \citep{Ken98} with some
scatter. The robustness of this technique still needs to be tested
with future observations, but it is encouraging that observers are
making progress towards a `direct' measurement of the SFR in DLAs.

On the other hand, there is a wealth of observational data on the
chemical abundances of DLAs from detailed studies of quasar absorption
line spectra (see Pettini 2003 for an excellent review). 
DLAs are generally inferred to be metal poor at all redshifts
\citep[e.g.][]{Pettini94, Lu96, Pettini99, Pro99}.  The column
density-weighted mean abundance of zinc is measured to be $\sim 1/13$
of solar ($\avg{[{\rm Zn/H}]}=-1.13$), which suggests that DLA
galaxies (i.e. galaxies in which DLAs arise) are at early stages of
their chemical evolution. The measured values of [Zn/H] span more 
than an order of magnitude at each redshift, and the chemical 
enrichment appears to proceed at different rates in different DLAs.  
The metallicity
distribution of DLAs peaks at a value in-between those of halo and
thick-disk stars of the Milky Way (Fig. 8 of Pettini 2003; Wyse \&
Gilmore 1995; Laird et al. 1988), suggesting that the kinematics of
DLAs should be closer to stellar haloes and bulges than to rotationally
supported disks. This finding is at odds with the proposal by
\citet{Pro98} that most DLAs are large disks with rotational
velocities in excess of 200 $\kms$.  An interesting fact is that there
is little evidence for any redshift evolution in the metallicity of
DLAs \citep{Pettini99, Pro00, Pro02}, despite the fact that most 
analytic models of global chemical evolution of the Universe predict
substantial evolution in metallicity as a function of time
\citep[e.g.][]{Pei95}.  However, most recently \citet{Pro03} have 
detected mild evolution with a significantly larger sample of DLAs. 
It is interesting to see if full cosmological simulations agree with 
observations or analytic models of global chemical evolution, and what 
they predict in detail for DLA metallicities.

Theoretical understanding of the physical properties of DLAs still
remains primitive. Several semi-analytic models of DLAs and their
chemical evolution exist \citep[e.g.][]{Pei95, Kau96, Jim99, Mo99,
Pra00, Som01, Maller01, Maller02, Lanf03}, but they often neglect the effect
of galaxy mergers, infall of gas, outflows due to supernovae, and the
interaction with the intergalactic medium (IGM). These dynamical
processes are expected to have significant influence on the chemical
evolution of DLAs, and it would therefore be ideal to study the
chemical evolution of DLAs with full cosmological hydro-simulations
which treat all of these effects dynamically and self-consistently.
The earlier numerical studies of Gardner et al. (1997a,b; 2001) for
example did not address the chemical evolution of DLAs in detail.

\begin{table*}
\begin{center}
\begin{tabular}{cccccccc}
\hline
Run & Boxsize & ${N_{\rm p}}$ & $m_{\rm DM}$ & $m_{\rm gas}$ &
$\epsilon$ & $z_{\rm end}$ & wind \\
\hline
\hline
R3  & 3.375 & $2\times 144^3$ &  $9.29\times 10^5$ & $1.43\times 10^5$ &0.94 & 4.00 & strong \cr
R4  & 3.375 & $2\times 216^3$ &  $2.75\times 10^5$ & $4.24\times 10^4$ &0.63 & 4.00 & strong \cr
\hline
O3  & 10.00 & $2\times 144^3$ &  $2.42\times 10^7$ & $3.72\times 10^6$ &2.78 & 2.75 & none \cr
P3  & 10.00 & $2\times 144^3$ &  $2.42\times 10^7$ & $3.72\times 10^6$ &2.78 & 2.75 & weak \cr
Q3  & 10.00 & $2\times 144^3$ &  $2.42\times 10^7$ & $3.72\times 10^6$ &2.78 & 2.75 & strong \cr
Q4  & 10.00 & $2\times 216^3$ &  $7.16\times 10^6$ & $1.10\times 10^6$ &1.85 & 2.75 & strong \cr
Q5  & 10.00 & $2\times 324^3$ &  $2.12\times 10^6$ & $3.26\times 10^5$ &1.23 & 2.75 & strong \cr
\hline						                        
D4  & 33.75 & $2\times 216^3$ &  $2.75\times 10^8$ & $4.24\times 10^7$ &6.25 & 1.00 & strong \cr
D5  & 33.75 & $2\times 324^3$ &  $8.15\times 10^7$ & $1.26\times 10^7$ &4.17 & 1.00 & strong \cr
\hline						                        
G4  & 100.0 & $2\times 216^3$ &  $7.16\times 10^9$ & $1.10\times 10^9$ &12.0 & 0.00 & strong \cr
G5  & 100.0 & $2\times 324^3$ &  $2.12\times 10^9$ & $3.26\times 10^8$ &8.00 & 0.00 & strong \cr
\hline
\end{tabular}
\caption{Simulations employed in this study.
The box-size is given in units of $\himpc$, ${N_{\rm p}}$ is the
particle number of dark matter and gas (hence $\times\, 2$), $m_{\rm
DM}$ and $m_{\rm gas}$ are the masses of dark matter and gas
particles in units of $\himsun$, respectively, $\epsilon$ is the
comoving gravitational softening length in units of $\hikpc$,
and $z_{\rm end}$ is the ending redshift of the simulation. The
value of $\epsilon$ is a measure of spatial resolution.  From the 
top to the bottom row, we refer to R3 \& R4 collectively as
`R-series', the next 5 runs (O3 to Q5) are called `Q-series', D4 \& D5
are called `D-series', and G4 \& G5 are called `G-series'. The
`strong-wind' simulations form a subset of the runs analysed by
\citet{SH03b}.
\label{table:sim}}
\end{center}
\end{table*}

In our previous paper \citep{Nag03}, we studied the abundance of DLAs
using a series of cosmological SPH simulations with varying resolution
and feedback strength. We showed that the $\Lam$CDM model is able to
account for the observed abundance of DLAs at redshift $z=3-4.5$ quite
well. Another important conclusion of our study was that earlier
numerical work overestimated the DLA abundance significantly due to
insufficient numerical resolution, lack of efficient feedback
processes, and inaccuracies introduced in cooling processes when 
conventional formulations of SPH are used, giving rise to an overcooling
problem.  In our simulation methodology, we avoided these numerical
difficulties by using a novel version of SPH \citep{SH02} that is
based on integrating the entropy as an independent thermodynamic
variable \citep[e.g.][]{Lucy77, Her93}, and which takes variations of
the SPH smoothing lengths self-consistently into account.

In this paper, we focus on the overall distribution of the star
formation rate and metallicity of DLAs in the redshift range
$z=0-4.5$.  We are interested in how these quantities appear in our
new SPH simulations, in light of our refined modelling of star
formation and feedback by galactic winds, and our new formulation of
SPH. In particular, we will examine the dependence of our results on
feedback strength, and also investigate the influence of numerical
resolution on the physical properties of DLAs.

The evolution of metallicity in DLAs has previously been studied with
a cosmological hydro-simulation by \citet{Cen02}, using an Eulerian
code with a fixed mesh-size.  The methodology of this simulation is
very different from ours, but it is clearly interesting to compare their
results to what we find. We discuss this comparison in
Section~\ref{section:discussion}.

The paper is organised as follows.  In
Section~\ref{section:simulation}, we briefly describe the novel
features of our simulations, putting particular emphasis on the star
formation model and the treatment of metal production.  We show a few
examples of visual representations of haloes and associated physical
quantities in Section~\ref{section:visual}.  We then discuss the
distribution of \HI column density as a function of halo mass in
Section~\ref{section:column_mass}, projected stellar mass density in
Section~\ref{section:starcolumn}, projected SFR density in
Section~\ref{section:sfrcolumn}, and projected metallicity of DLAs in
Section~\ref{section:metalcolumn}.  The evolution of the mean
metallicity is described in Section~\ref{section:mean-metal}. Finally,
we discuss the implications of our work in
Section~\ref{section:discussion}.


\section{Simulations}
\label{section:simulation}

We analyse a large set of cosmological SPH simulations that differ in
box size, mass resolution and feedback strength, as summarised in
Table \ref{table:sim}. In particular, we consider box sizes ranging
from 3.375 to $100\himpc$ on a side, with particle numbers
between $2\times 144^3$ and $2\times 324^3$, allowing us to probe
gaseous mass resolutions in the range $4.2 \times 10^4$ to $1.1\times
10^9\,h^{-1}{\rm M}_\odot$.  These simulations are partly taken from a
study of the cosmic star formation history by \citet{SH03b},
supplemented by additional runs with weaker or no galactic winds. The
joint analysis of this series of simulations allows us to
significantly broaden the range of spatial and mass-scales that we can
probe compared to what is presently attainable within a single
simulation.

There are three main new features of our simulations.  First, we use a
``conservative entropy'' formulation of SPH \citep{SH02}.  This
approach has several advantages over conventional versions of SPH.
Since the energy equation is written with the entropy as the
independent thermodynamic variable, rather than the thermal energy,
the `p\,dV' term is not evaluated explicitly, reducing noise from
smoothed estimates of e.g. the density.  By including terms involving
derivatives of the density with respect to the particle smoothing
lengths, this approach explicitly conserves entropy (in regions
without shocks), as well as momentum and energy, even when smoothing
lengths evolve adaptively, avoiding the problems noted by
e.g. Hernquist (1993).  Furthermore, this formulation moderates the
overcooling problem present in earlier formulations of SPH \citep[see
also][]{Yoshida02, Pearce99, Croft01}.

Second, we employ an effective sub-resolution model described by
\citet{SH03a} to treat star formation and its regulation by supernovae
feedback in the dense interstellar medium (ISM).  In this model, the
highly overdense gas of the ISM is pictured to be a two-phase fluid
consisting of cold clouds in pressure equilibrium with a hot ambient
phase. Each gas particle represents a statistical mixture of these
phases. Cold clouds grow by radiative cooling out of the hot medium,
and this material forms the reservoir of baryons available for star
formation.  We assume that star formation takes place on a
characteristic time-scale $\tstar$, and that a mass fraction $\beta$
of these stars are short-lived and instantly die as supernovae so 
that the local SFR is: \beq
\frac{\dd\rhostar}{{\dd t}} = (1-\beta)\frac{\rho_c}{\tstar}, \eeq
where $\rho_c$ is the gas density of the cold cloud phase.  The value
of $\beta$ is determined by the initial mass function of the stellar
population, and here we adopt $\beta=0.1$.  The star formation
time-scale $\tstar$ is taken to be proportional to the local dynamical
time of the gas: \beq \tstar(\rho) =
\tzstar\rp{\frac{\rho}{\rhoth}}^{-1/2}, \eeq where the value of
$\tzstar=2.1 ~\Gyr$ is chosen to match the Kennicutt law
(1998). Because of its density dependence, the star formation
time-scale becomes short in very high density regions, approximately
modelling intense starburst activity.  In our model, this parameter
simultaneously determines a threshold density $\rhoth$, above which
the multiphase structure in the gas, and hence star formation, is
allowed to develop.  This physical density is $8.6\times 10^6
h^2\Msun\kpc^{-3}$ for all simulations in this study, corresponding to
a comoving baryonic overdensity of $7.7\times 10^5$ at $z=0$. See
\citet{SH03a} for a description for how $\rhoth$ is determined
self-consistently within the model.

In our simulation code, the mass of a new star particle is fixed to
$m_{\star} = m_0/N_g$, where $m_0$ is the initial mass of each gas
particle, and $N_g$ (which we take to be 2 here) is the number of
``generations'' of stars each gas particle may spawn.  New star
particles are created probabilistically with expectation value
consistent with the current SFR of each gas particle.

Once star formation occurs, the resulting supernova explosions are
assumed to deposit energy into the hot gas with an average of $\esn =
4\times 10^{48} \ergs~\Msun^{-1}$ for each solar mass of stars
formed, or 0.4\% of the canonical SN energy of $10^{51}~\ergs$.
 The efficiency of SN energy feedback is often described 
as the ratio of the returned SN energy to the rest mass energy of 
the star formed, in which case our SN feedback efficiency translates to 
$\esn/\Msun c^2 = 2.2\times 10^{-6}$. The heating rate due to SN is 
\beq \left.\frac{\dd}{\dd t}(\rho_h u_h)\right|_{\rm SN} = 
\esn \frac{\dd\rhostar}{\dd t} =
\beta \usn \frac{\rho_c}{\tstar}, \eeq where $\usn \equiv
(1-\beta)\beta^{-1}\esn$ is the `SN energy scale' corresponding to a
`SN temperature' of $T_{\rm SN}=2\mu\usn/3k \simeq 10^8\,{\rm K}$, and
$\rho_h$ and $u_h$ are the density and thermal energy of the hot phase
of the gas, respectively.

We assume that SN explosions also evaporate the cold clouds, and transfer
cold gas back into the ambient hot phase. This can be described as
\beq
\left.\frac{\dd \rho_c}{\dd t}\right|_{\rm EV} = A\beta\frac{\rho_c}{\tstar}.
\eeq
The efficiency parameter $A$ has the density dependence
\beq
A(\rho) = A_0\rp{\frac{\rho}{\rhoth}}^{-4/5},
\eeq
following \citet{McKee77}. The value of $A_0$ is fixed to 1000
by requiring $T_{\rm SN}/A_0 = 10^5$ K.
This evaporation process of cold clouds establishes a tight 
self-regulation mechanism for star formation in the ISM,
where the ambient hot medium quickly evolves towards an 
equilibrium temperature equal to
\beq
u_h = \frac{\usn}{A+1} + u_c
\label{eq:hot-temp}
\eeq in a time-scale shorter than the star formation time-scale.
Therefore, it is a good approximation to assume that the condition of
self-regulated star formation always holds. We can then avoid treating
the mass exchange between the hot and cold phases explicitly.  This
simplifies the code, makes it faster, and reduces its memory
consumption.  In this `effective' model, which is adopted for all the
runs used in the present study, the thermal energy of the hot phase is
simply given by equation~(\ref{eq:hot-temp}), and that of the cold
phase is fixed to the equivalent of $T_c = 1000~\K$. Our result does
not depend on the choice of $T_c$ as long as $T_c \ll 10^4~\K$,
because the current code does not include molecular hydrogen 
or metal line cooling
which would act at temperatures below $10^4~\K$. The mass fraction in
the cold clouds for the particles in the multiphase medium can be computed by
\beq x \equiv \frac{\rho_c}{\rho} =
1+\frac{1}{2y}-\sqrt{\frac{1}{y}+\frac{1}{4y^2}},
\label{eq:x}
\eeq
where
\beq
y \equiv \frac{\tstar\Lambda_{\rm net}(\rho,u_h)}{\rho[\beta\usn-(1-\beta)u_c]}.
\eeq
Here, $\Lambda_{\rm net}(\rho,u)$ is the usual cooling function.

If the physical density of a gas particle is lower than $\rhoth$, 
the particle represents ordinary gas in a single phase. In this case, 
the neutral hydrogen mass of the particle can be computed as
\beq 
m_{\rm HI} = \Nh \cdot \XH \cdot m_{\rm gas} \quad (\rho<\rho_{\rm th}),
\eeq
where $\XH=0.76$ is the primordial mass fraction of hydrogen, and
$\Nh$ is the number density of neutral hydrogen atoms in units
of the total number density of hydrogen nuclei. The quantity $\Nh$ is 
computed by solving the ionisation balance as a function of 
density and current UV background flux.
If the gas density is higher than $\rhoth$, the gas particle represents 
multiphase gas, such that the mass of neutral hydrogen can be computed as
\beq
m_{\rm HI} = x \cdot \XH \cdot m_{\rm gas} \quad (\rho>\rho_{\rm th}),
\eeq
where $x$ is the mass fraction of cold clouds given by equation~(\ref{eq:x}).

The simulation also keeps track of metal enrichment, and the dynamical
transport of metals by the motion of the gas particles.  Metals are
produced by stars and returned into the gas by SN explosions. The mass
of metals returned is $\Delta M_Z = y_\star \Delta M_{\star}$, where
$y_\star=0.02$ is the yield, and $\Delta M_{\star}$ gives the mass of
newly formed long-lived stars. Assuming that each gas element behaves
as a closed box locally, with metals being instantaneously mixed
between cold clouds and the ambient hot gas, the metallicity
$Z=M_Z/M_g$ of a star-forming gas particle increases during one
timestep $\Delta t$ by \beq \Delta Z = (1-\beta)y_\star
\,x\,\frac{\Delta t}{\tstar}.  \eeq When a star particle is generated,
it simply inherits the metallicity of its parent gas particle.

\begin{figure*}
\begin{center}
\resizebox{7.3cm}{!}{\includegraphics{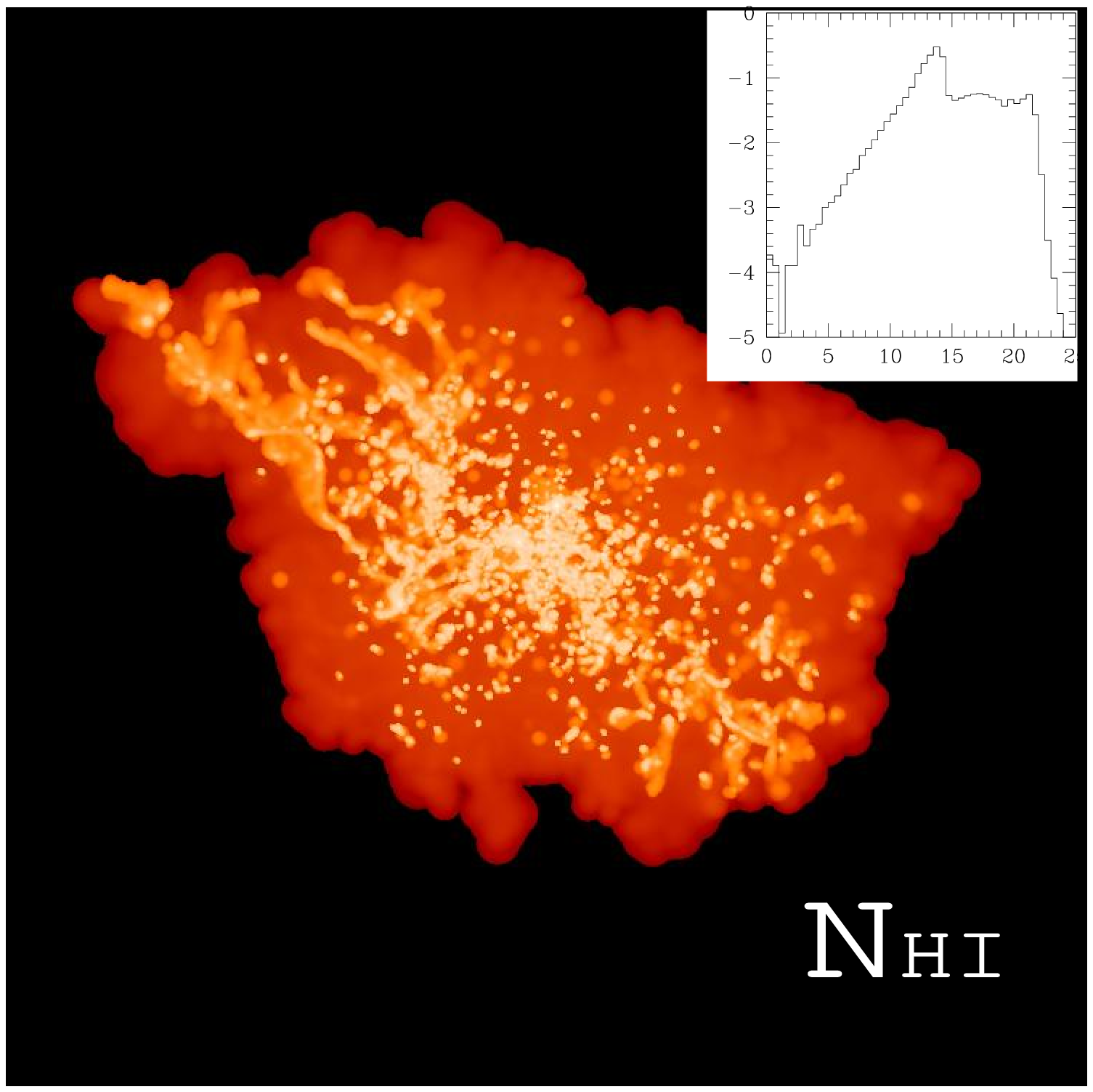}}%
\hspace{0.13cm}
\resizebox{7.3cm}{!}{\includegraphics{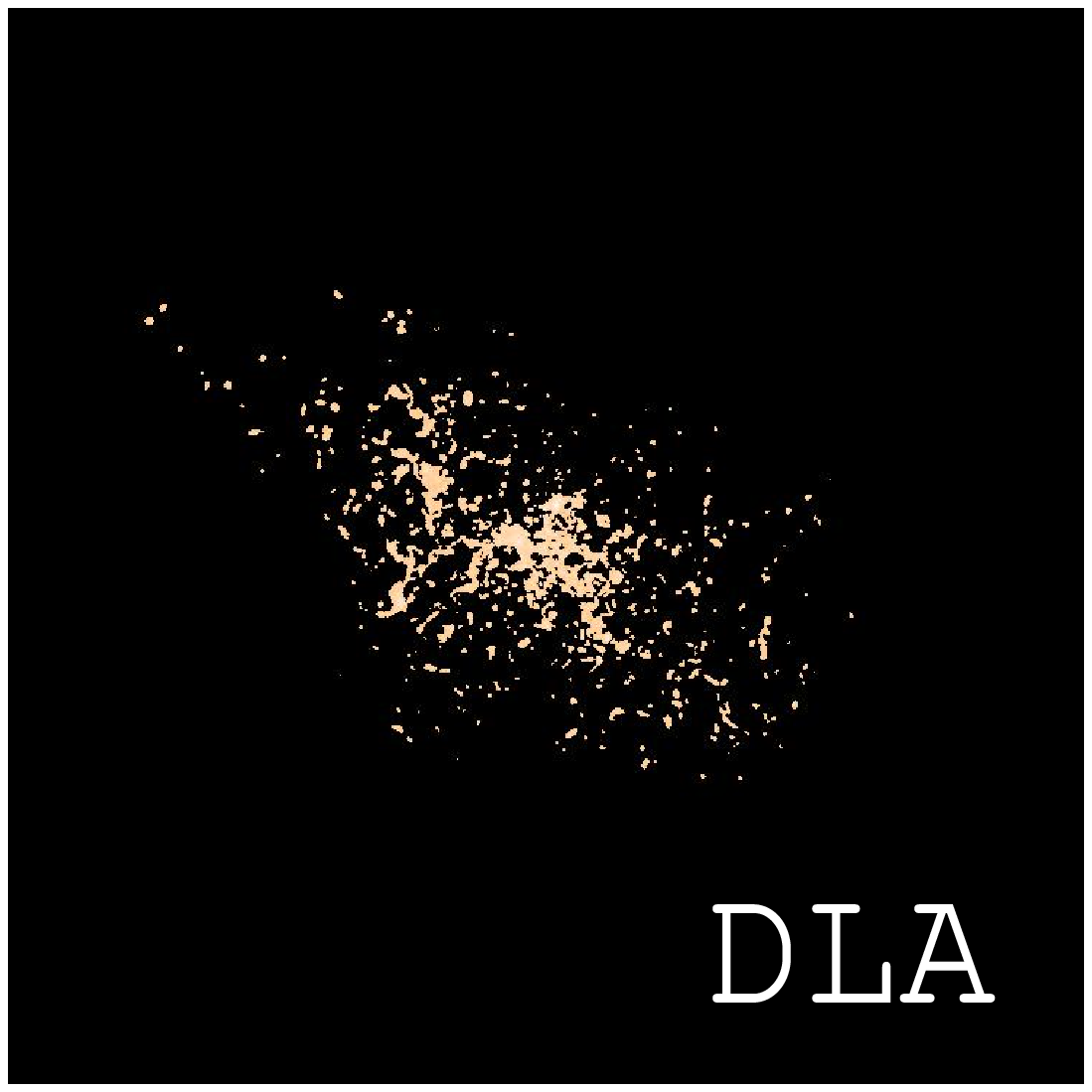}} \\%
\vspace{0.08cm}
\resizebox{7.3cm}{!}{\includegraphics{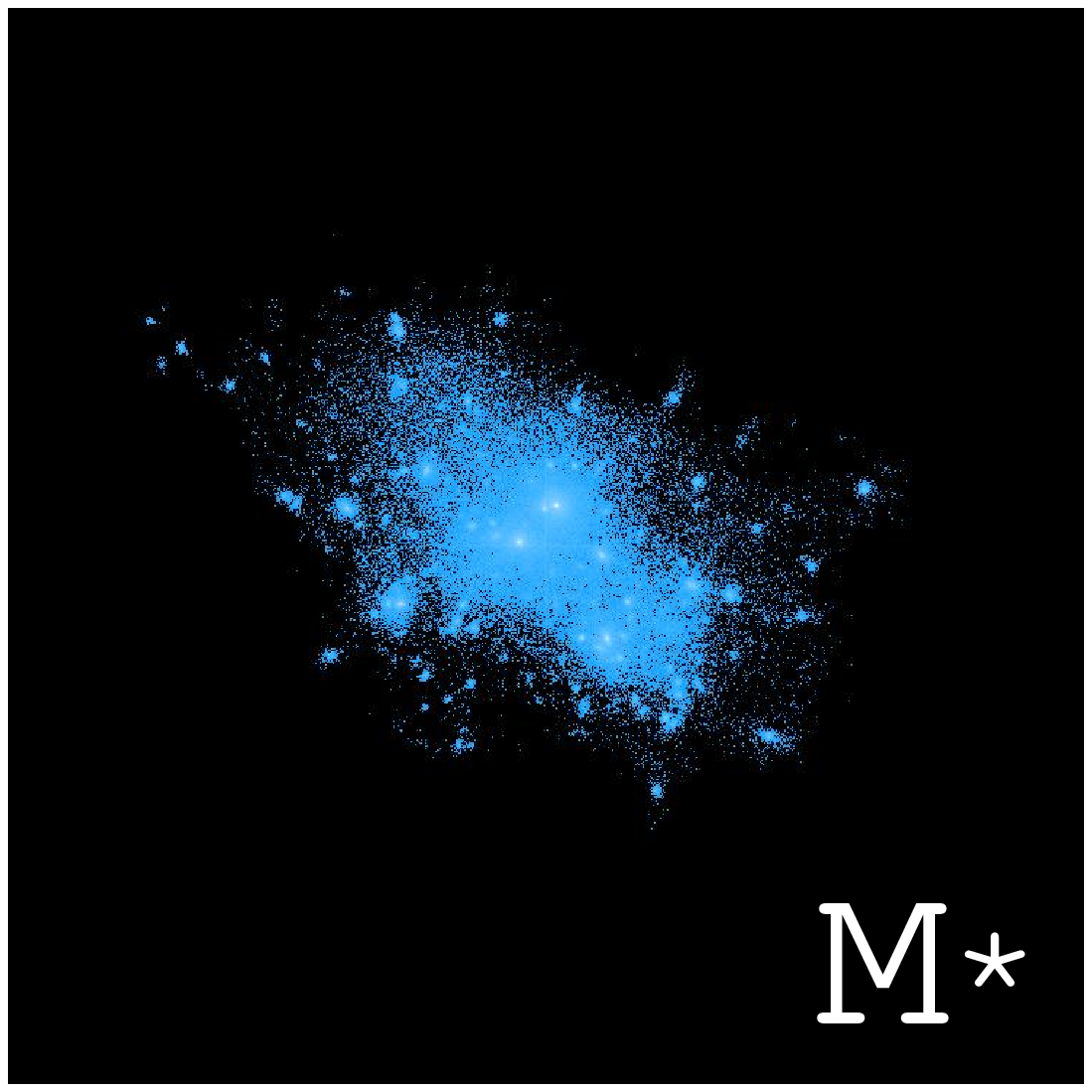}}%
\hspace{0.13cm}
\resizebox{7.3cm}{!}{\includegraphics{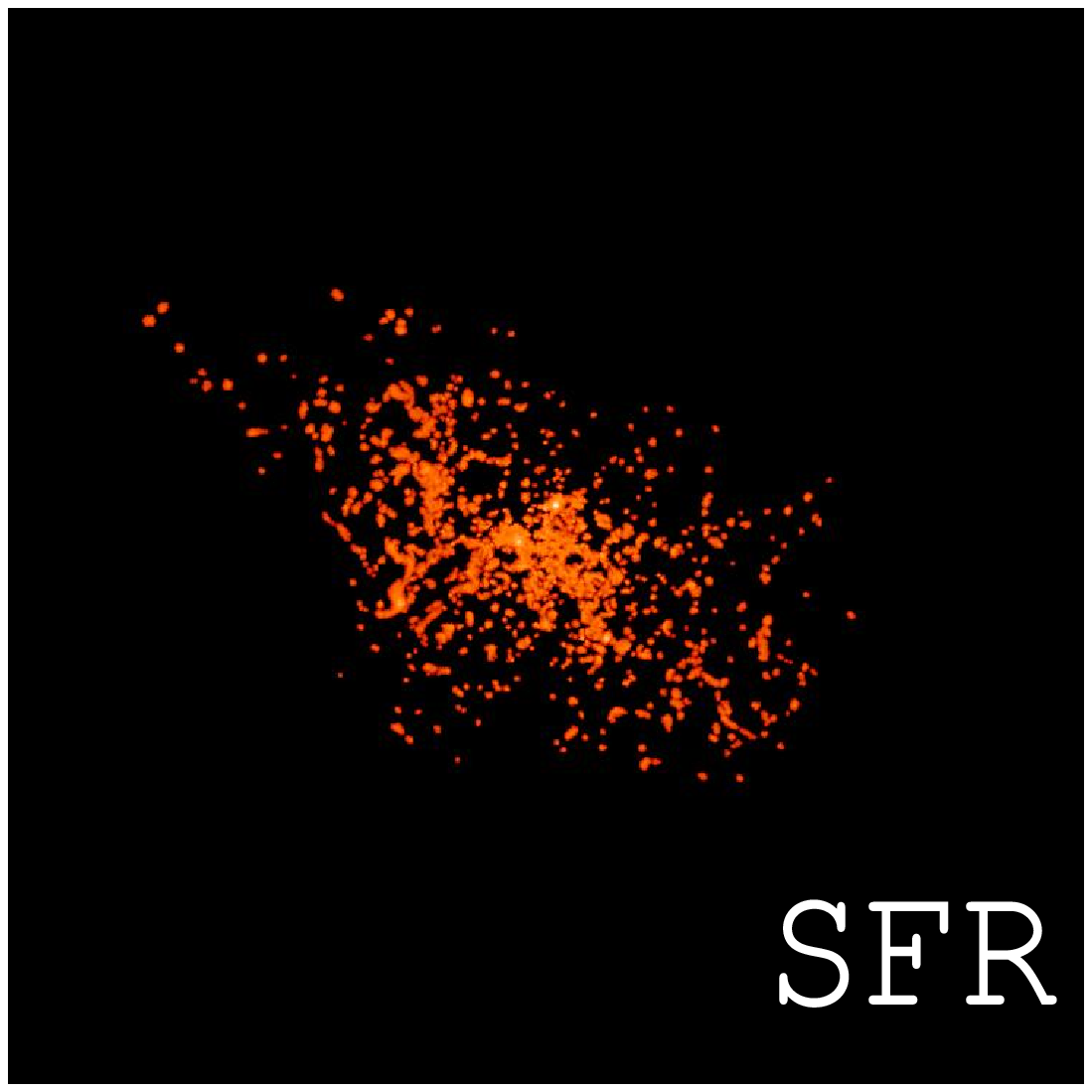}} \\%
\vspace{0.08cm}
\resizebox{7.3cm}{!}{\includegraphics{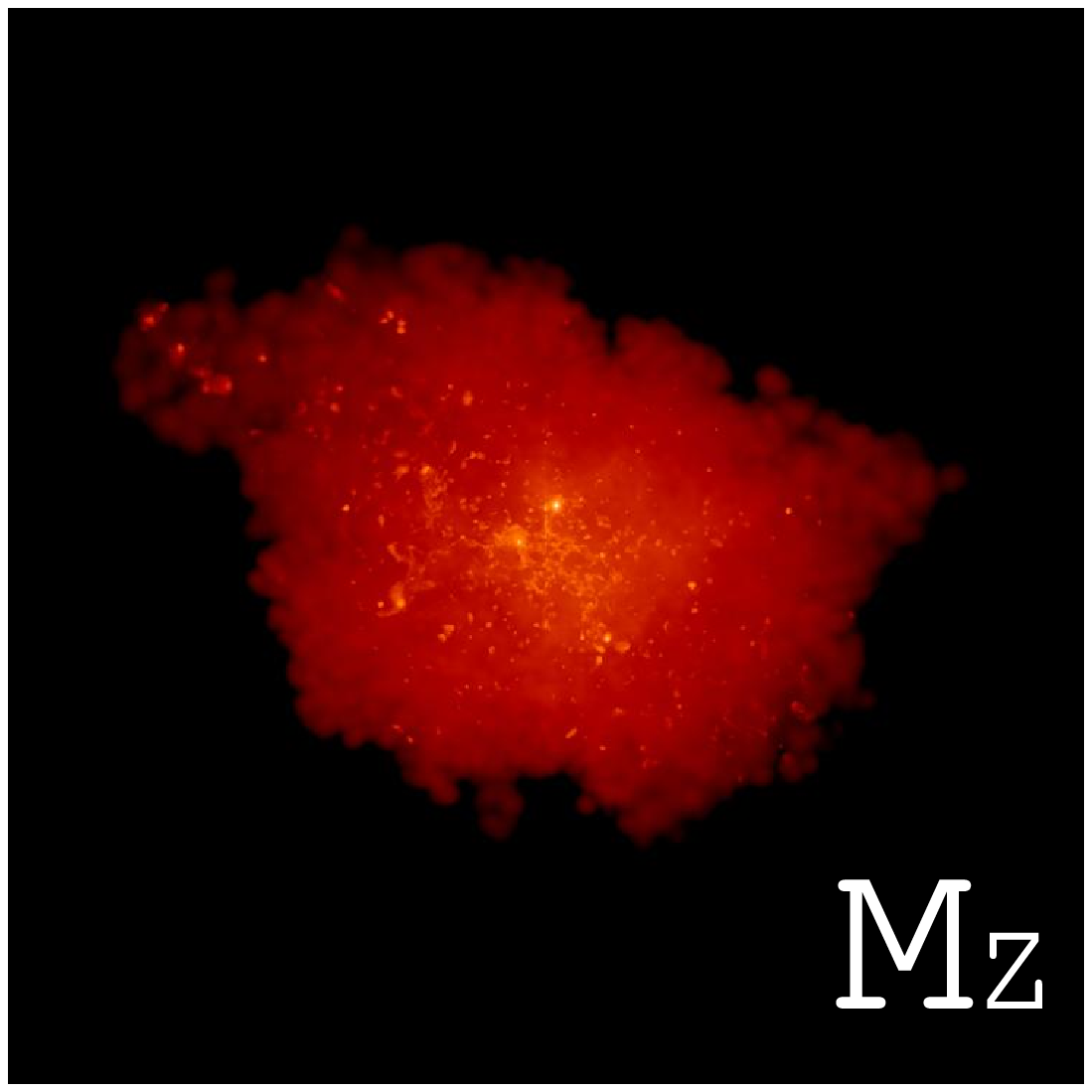}}%
\hspace{0.13cm}
\resizebox{7.3cm}{!}{\includegraphics{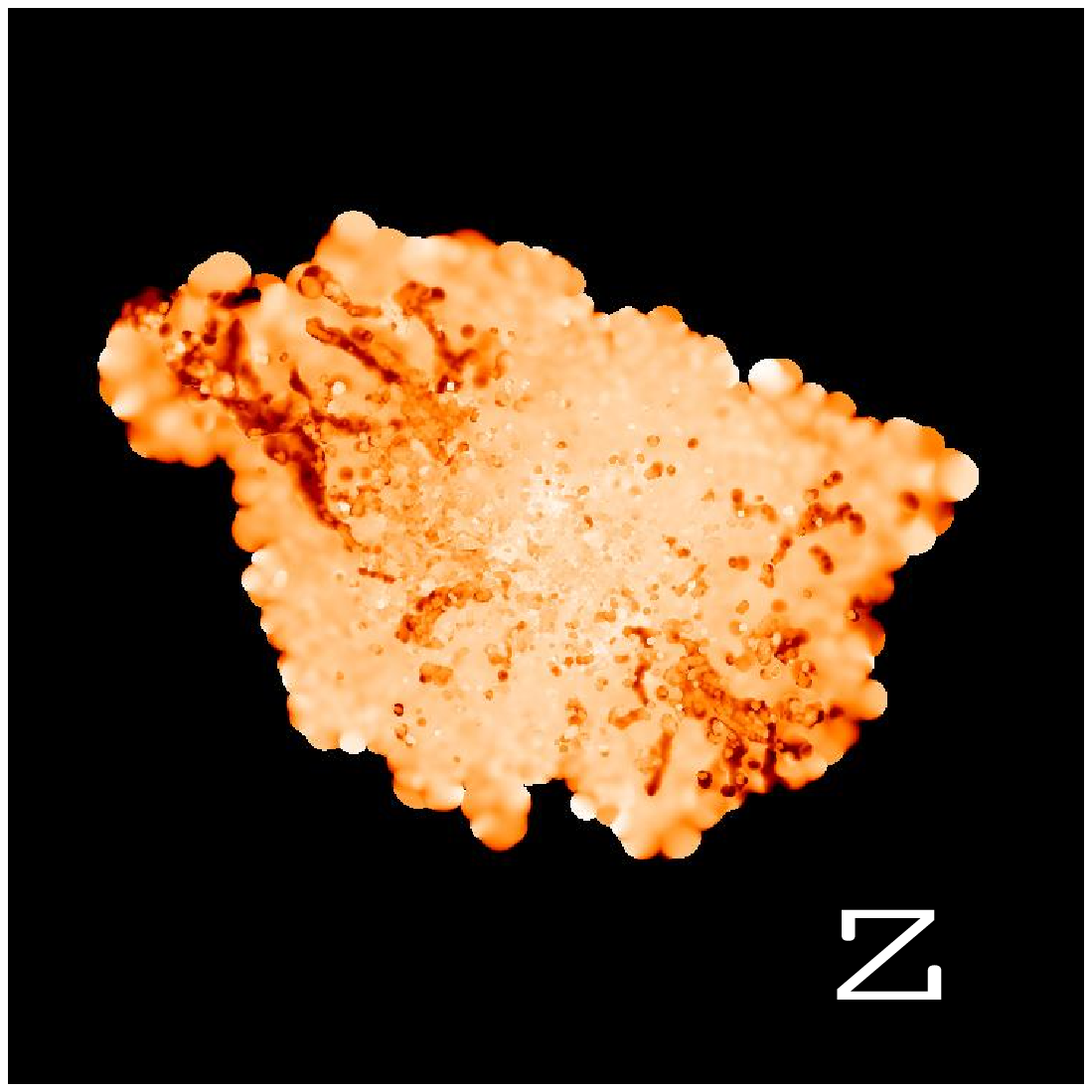}}\\%
\caption{Projected spatial distribution of various quantities (all in
log scale) for the most massive halo of mass $\Mhalo=1.7\times
10^{12}\hinv\Msun$ at $z=3$ in the `Q5'-run.  From top left to bottom
right: $\NHI$, DLAs ($\log\NHI>20.3$), stellar surface mass density,
SFR surface density, metal mass surface density, gas metallicity. The
size of each panel is comoving $\pm 457 \hikpc$ from the centre of the
halo.  The inset in the top left panel shows the probability
distribution function of lines-of-sight (${\rm d}n/{\rm d}\log\NHI$)
for this halo as a function of $\log\NHI$. One can see that the
majority of the red region is occupied by lines of sight with
$11<\log\NHI<15$.
\label{dla-pic.eps}}
\end{center}
\end{figure*}

Finally, the third important feature of our simulations is a 
phenomenological model for galactic winds, which we introduced in
order to study the effect of outflows on DLAs, galaxies, and the
intergalactic medium (IGM).  In this model, gas particles are stochastically
driven out of the dense star-forming medium by assigning an extra
momentum in random directions, with a rate and magnitude chosen to
reproduce mass-loads and wind-speeds similar to those observed. 
The wind mass-loss rate $\Mwdot$ is assumed to be proportional to 
the star formation rate, and the wind carries a fixed fraction 
$\chi$ of SN energy:
\beq
\Mwdot = \eta \Msdot, 
\eeq
and
\beq
\frac{1}{2} \Mwdot v_w^2 = \chi \esn \Msdot.
\eeq
We adopted a fixed value of $\eta=2$, and explored two values of
$\chi$, a `weak wind model' with $\chi=0.25$, and a `strong wind
model' with $\chi=1.0$ . Solving for the wind velocity from the above
two equations, the wind models correspond to speeds of 
$v_w = 242~\kms$ and $484~\kms$, respectively. This wind energy is 
not included in $\esn$ discussed above, so in our simulations 
$(1+\chi)\esn$ is the total SN energy returned into the gas.

Most of our simulations employ the `strong' wind model,
but for the $10\,h^{-1}{\rm Mpc}$ box (runs O3, P3, Q3,
Q4, Q5; collectively called `Q-series') we also varied the wind
strength.  Therefore, this Q-series can be used to study both the
effect of numerical resolution and the consequences of feedback from
galactic winds.  The runs in the other simulation series then allow
the extension of the strong wind results to smaller scales
(`R-Series'), or to larger box-sizes and hence lower redshift (`D-'
and `G-Series').  Our naming convention is such that runs of the same
model (box-size and included physics) are designated with the same
letter, with an additional number specifying the particle resolution.

Our calculations include a uniform UV background radiation field of a
modified \citet{Haa96} spectrum, where reionisation takes place 
at $z\simeq 6$ \citep[see][]{Dave99}, as suggested by observations
\citep[e.g.][]{Becker01} and radiative transfer calculations of the
impact of the stellar sources in our simulations on the IGM
\citep[e.g.][]{Sok03}.  The radiative cooling and heating rate is
computed as described by \citet{Katz96}.  The adopted cosmological
parameters of all runs are $(\Om,\Ol,\Ob,\sigma_8, h)= (0.3, 0.7,
0.04, 0.9, 0.7)$. The simulations were performed on the Athlon-MP
cluster at the Center for Parallel Astrophysical Computing (CPAC) at
the Harvard-Smithsonian Center for Astrophysics, using a modified
version of the parallel {\small GADGET} code \citep{Gadget}.


\section{Visual representations}
\label{section:visual}

\begin{figure*}
\begin{center}
\resizebox{5.2cm}{!}{\includegraphics{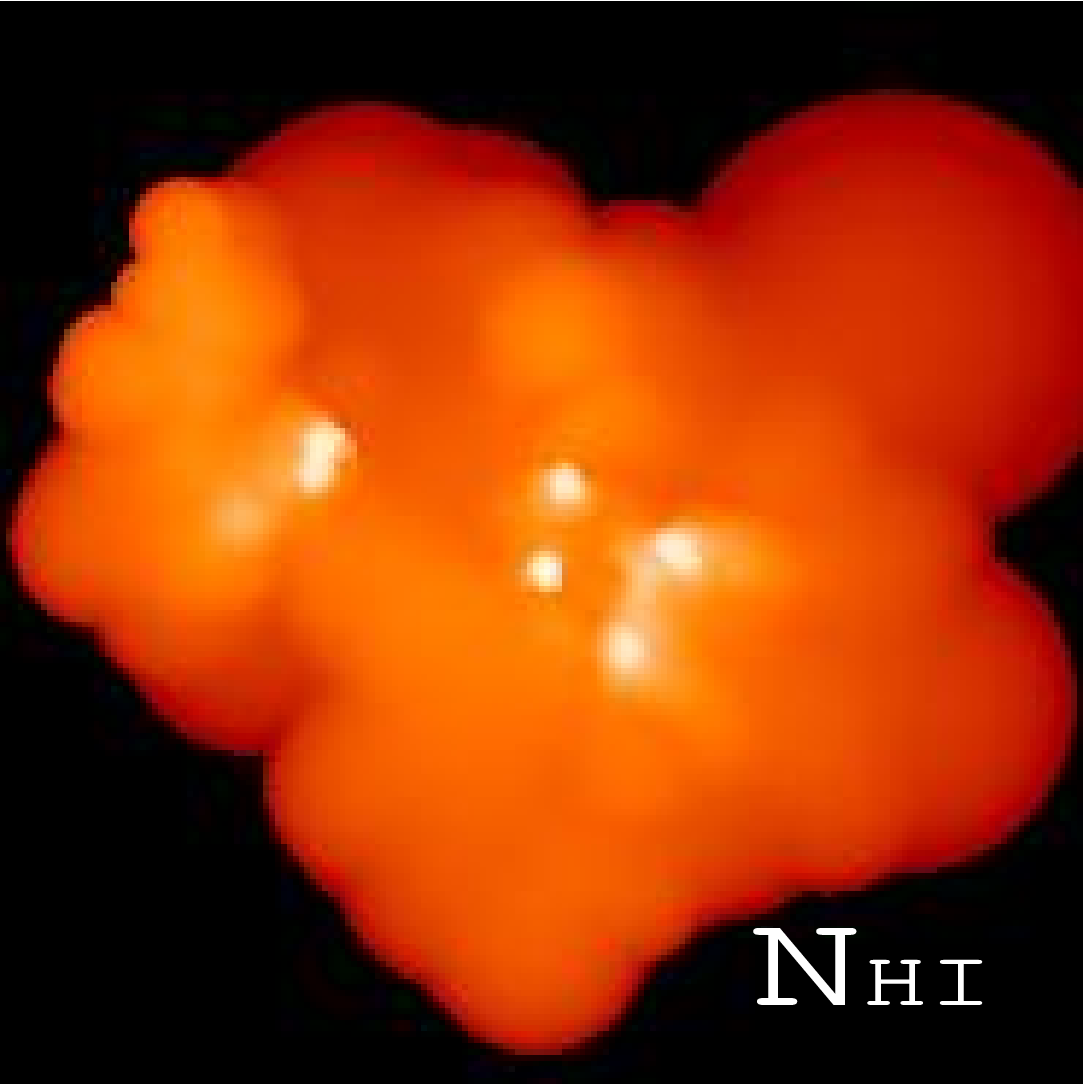}}%
\hspace{0.13cm}
\resizebox{5.2cm}{!}{\includegraphics{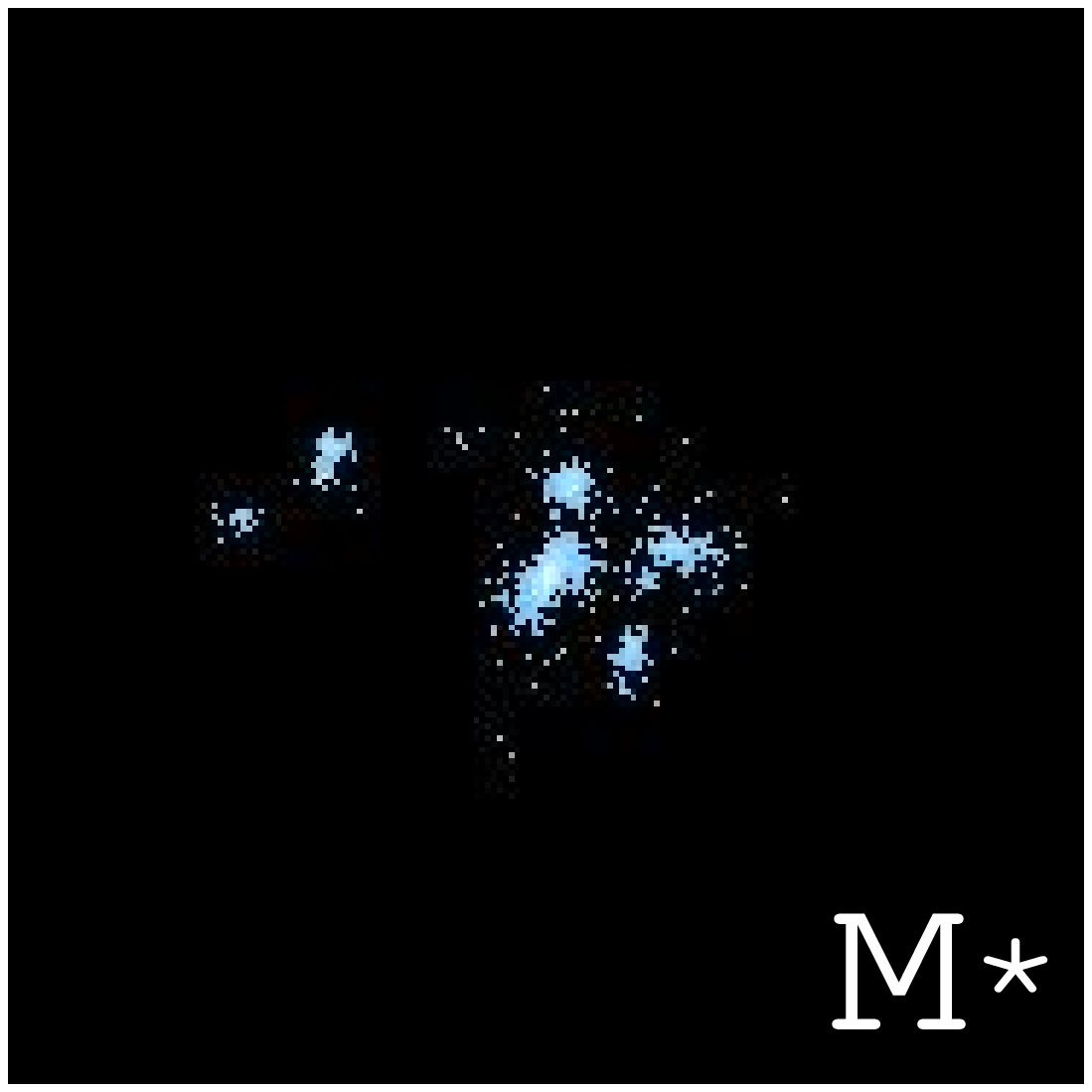}}%
\hspace{0.13cm}
\resizebox{5.2cm}{!}{\includegraphics{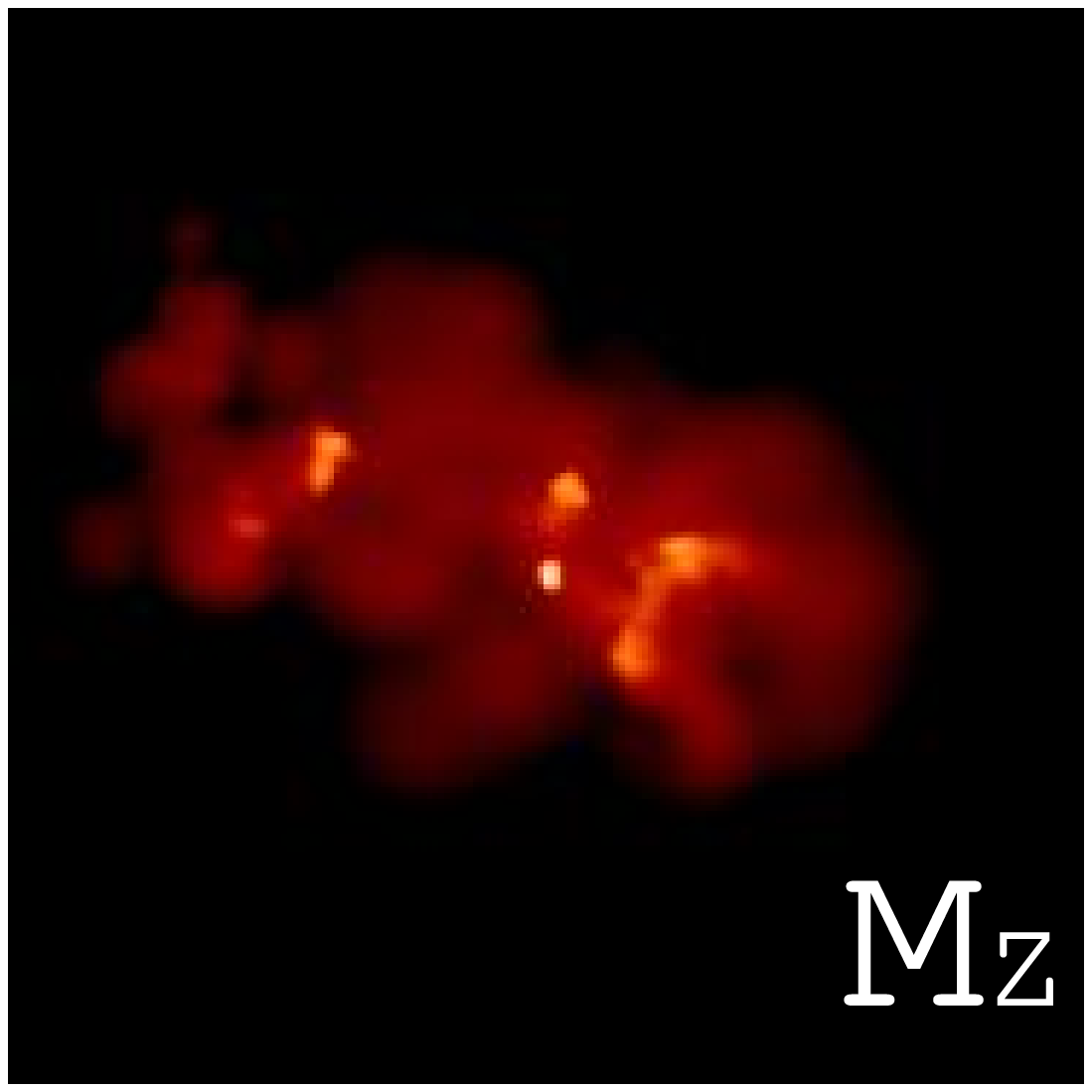}}\\%
\vspace{0.13cm}
\resizebox{5.2cm}{!}{\includegraphics{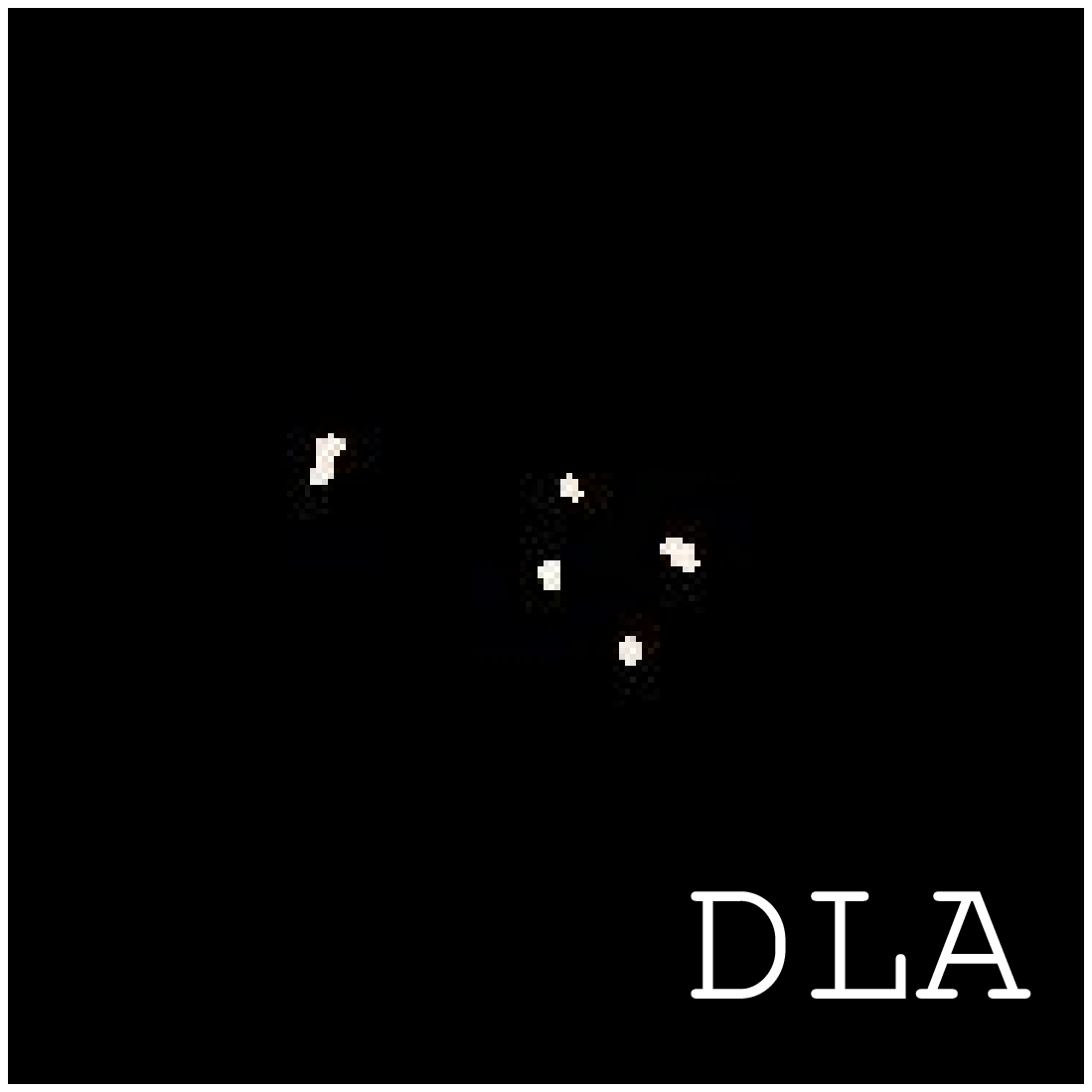}}%
\hspace{0.13cm}
\resizebox{5.2cm}{!}{\includegraphics{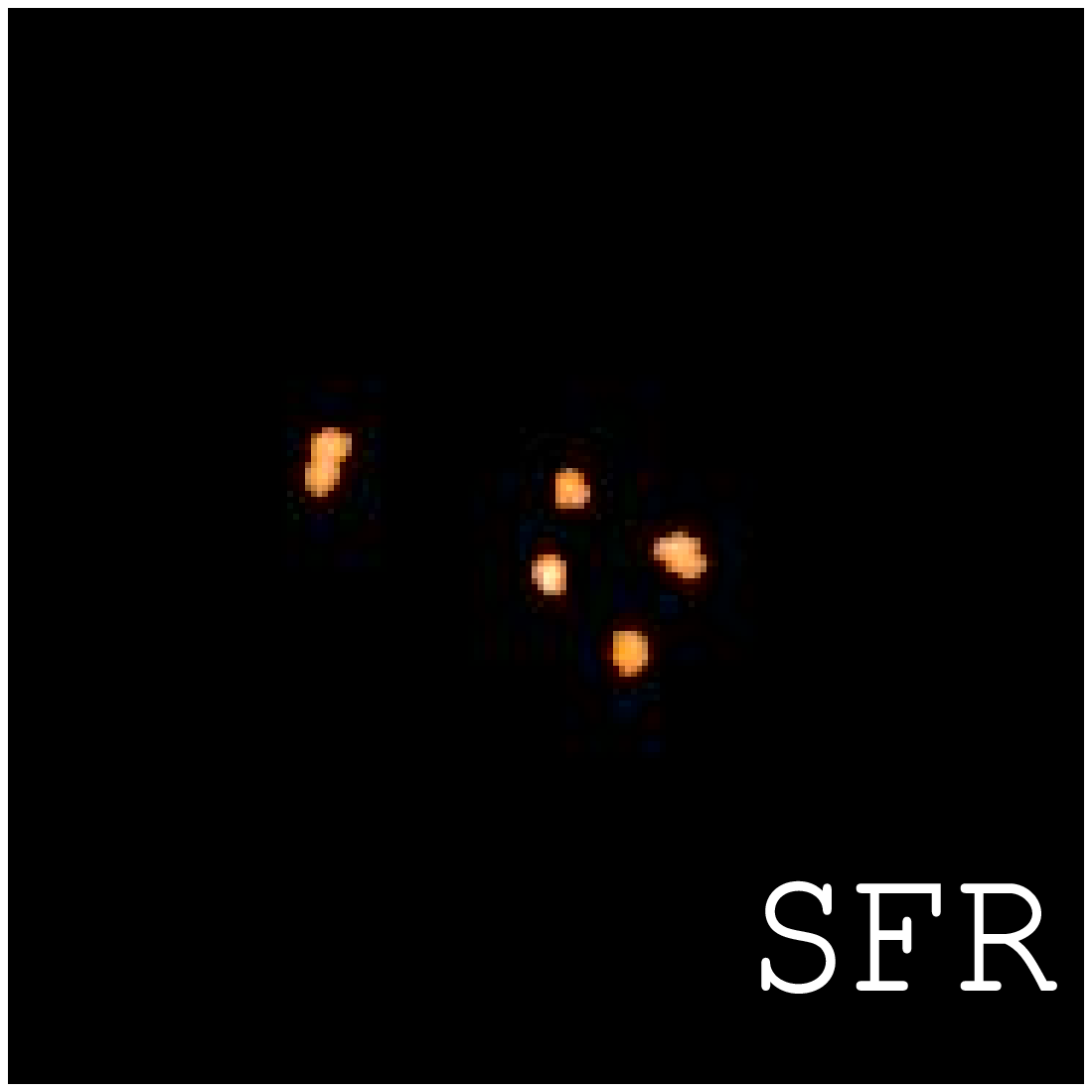}} %
\hspace{0.05cm}
\resizebox{5.2cm}{!}{\includegraphics{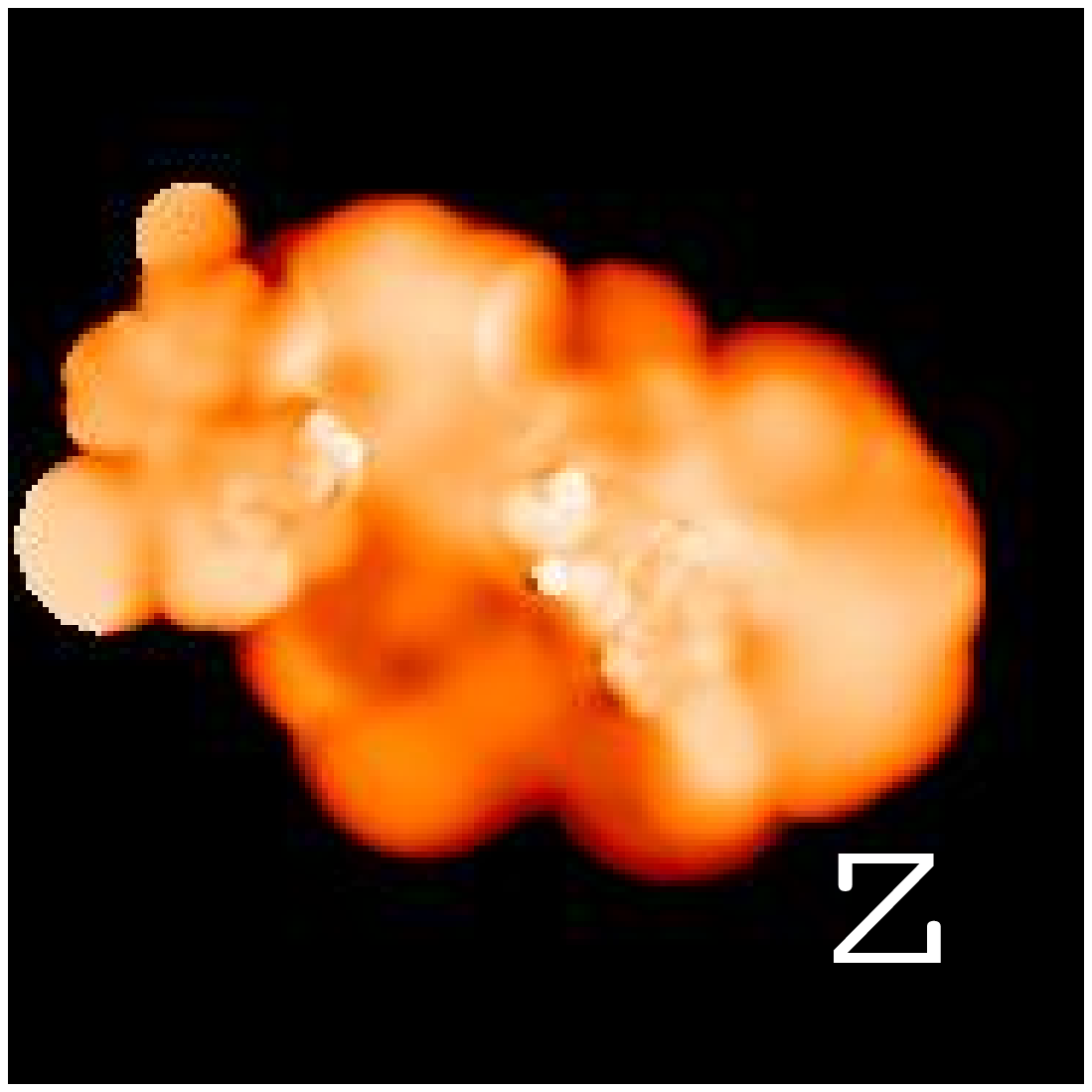}}\\%
\caption{Projected spatial distribution of various quantities for a
halo of mass $\Mhalo=2.6\times 10^{10}\hinv\Msun$ at $z=3$ in the `Q5'-run.
Left column: $\NHI$ (top) and DLAs (bottom).  Middle column: stellar
surface mass density (top) and SFR surface density (bottom).  Right
column: metal mass surface density (top) and gas metallicity (bottom).
The size of each panel is comoving $\pm 112 \hikpc$ from the centre of
the halo.
\label{dla-pic2.eps}}
\end{center}
\end{figure*}

\begin{figure*}
\begin{center}
\resizebox{4cm}{!}{\includegraphics{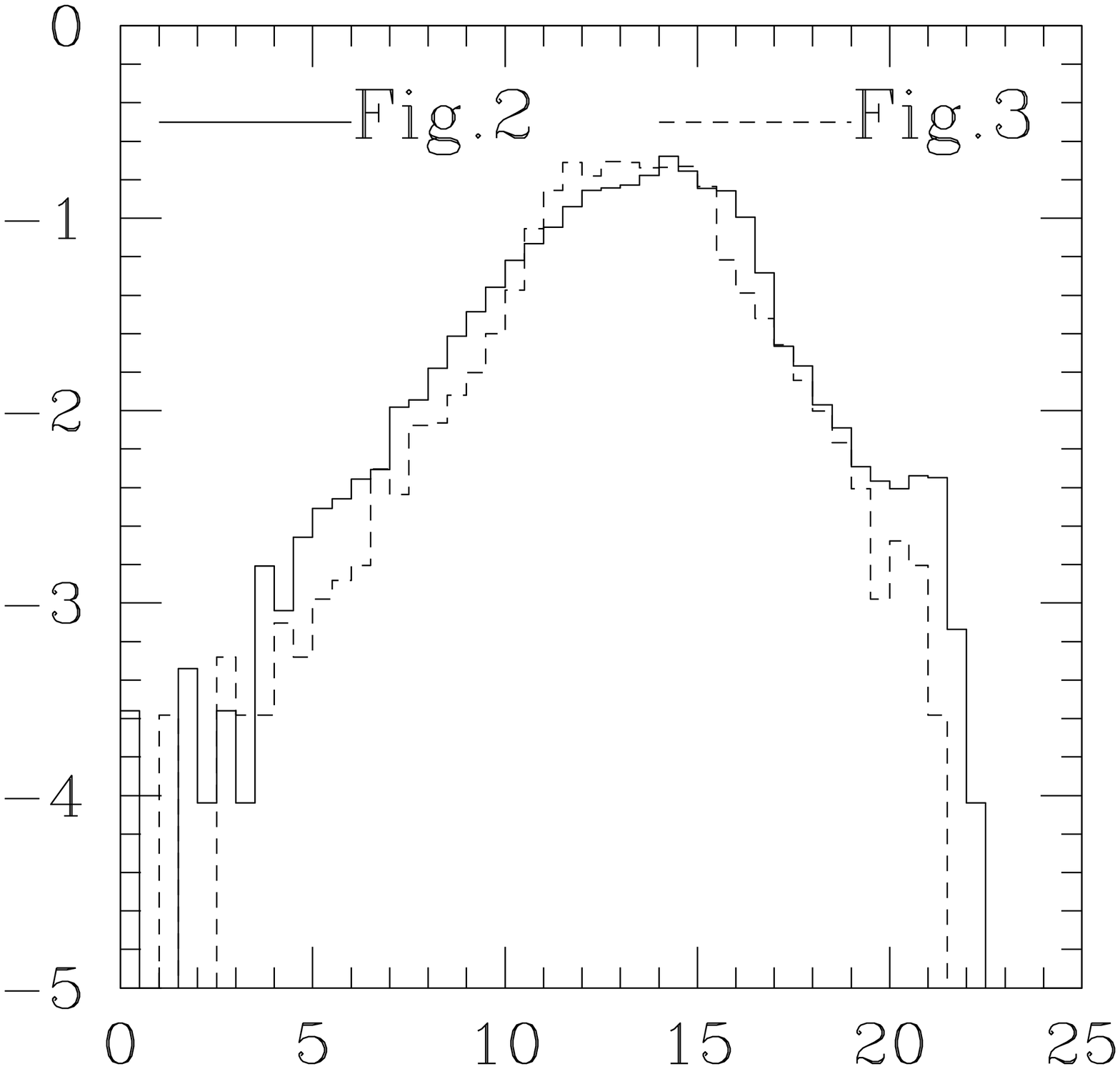}}%
\hspace{0.13cm}
\resizebox{3.8cm}{!}{\includegraphics{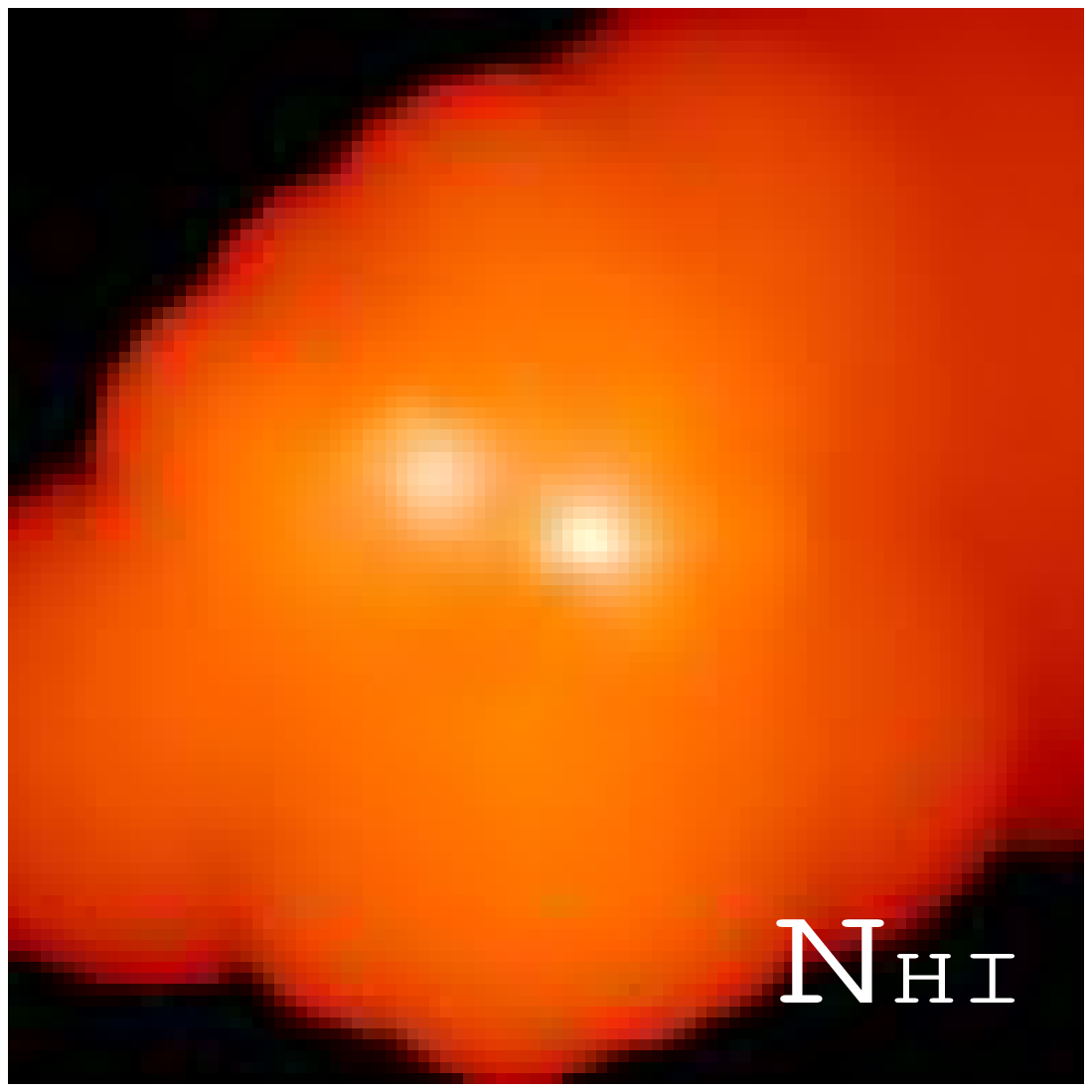}}%
\hspace{0.13cm}
\resizebox{3.8cm}{!}{\includegraphics{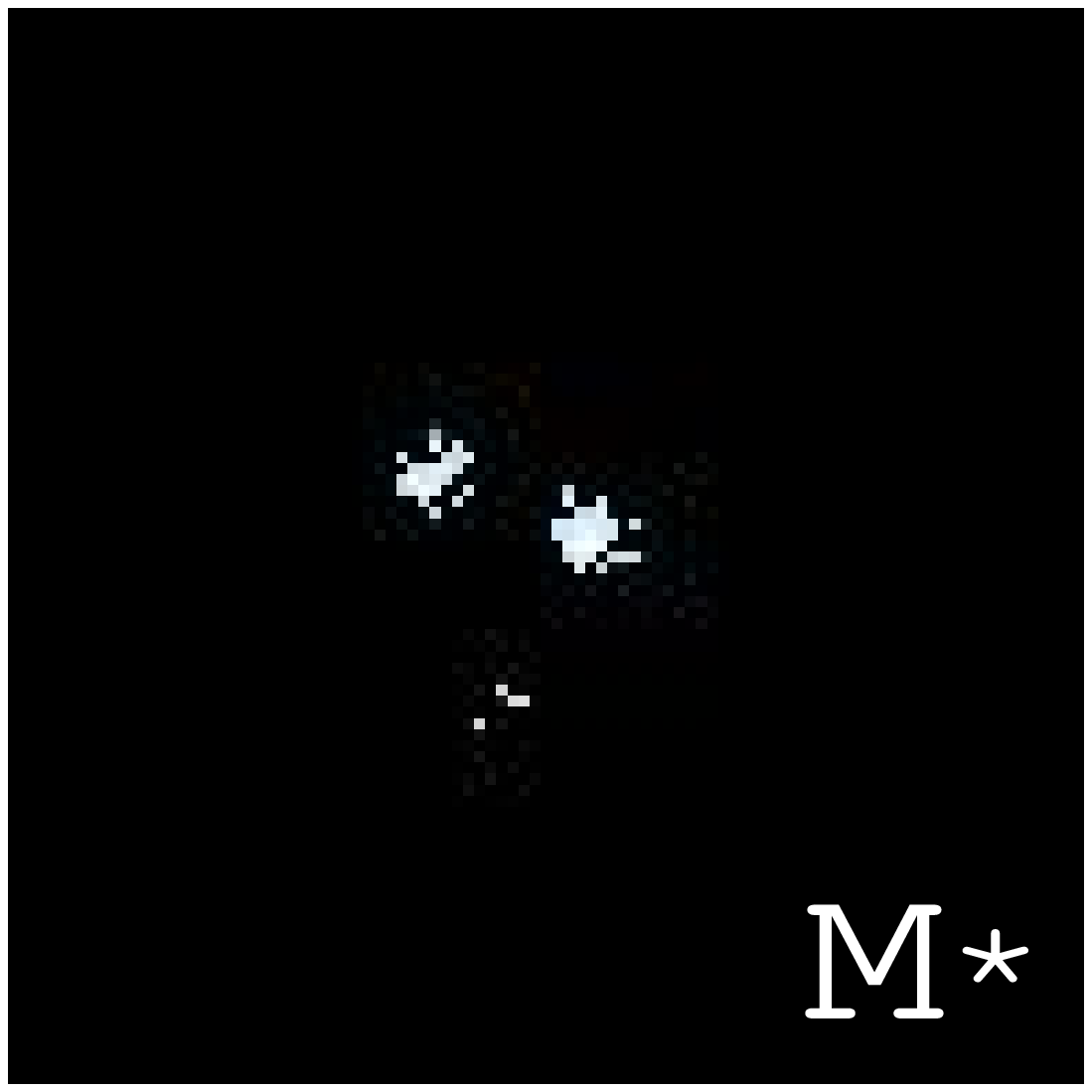}}%
\hspace{0.13cm}
\resizebox{3.8cm}{!}{\includegraphics{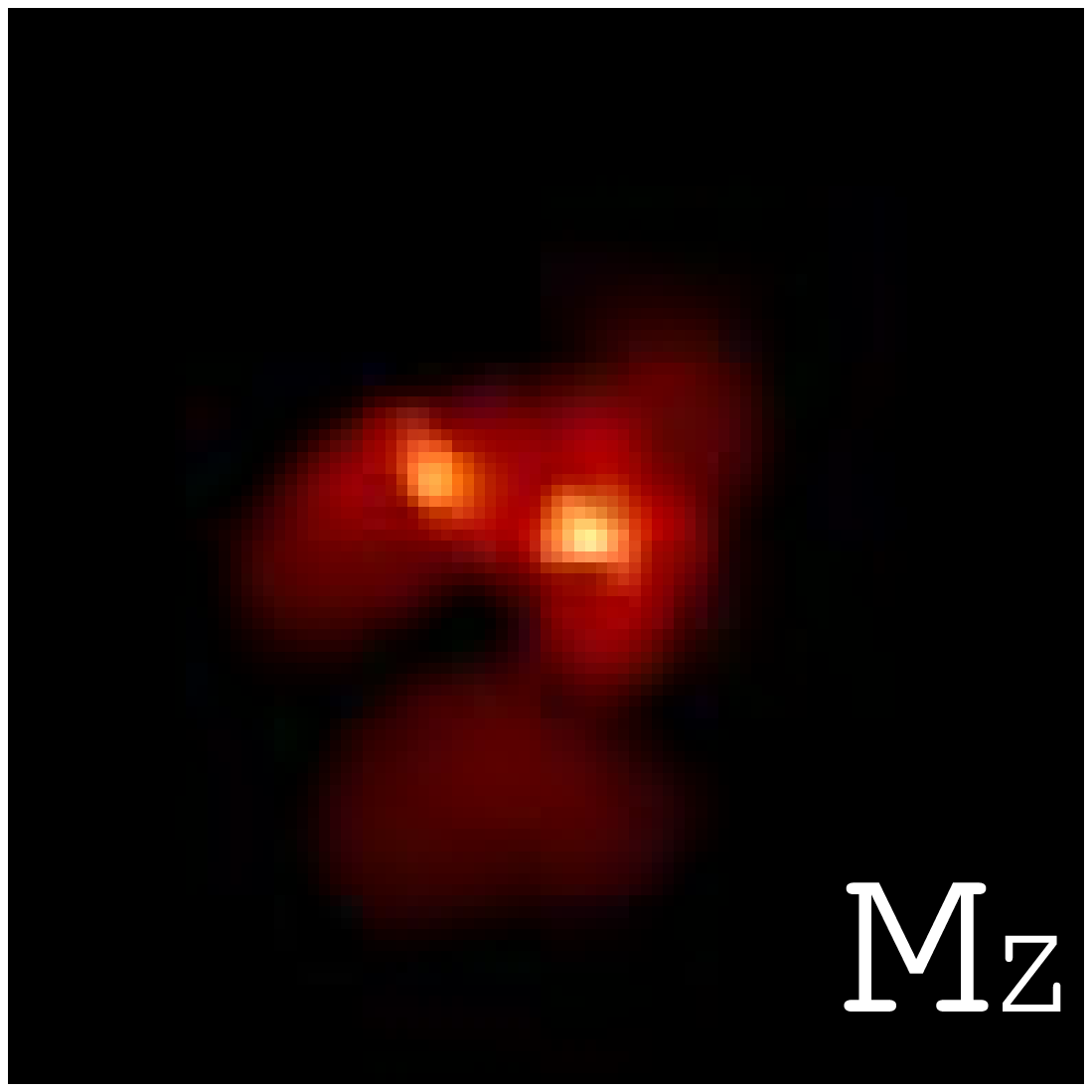}}\\%
\vspace{0.13cm}
\hspace{4.13cm}
\resizebox{3.8cm}{!}{\includegraphics{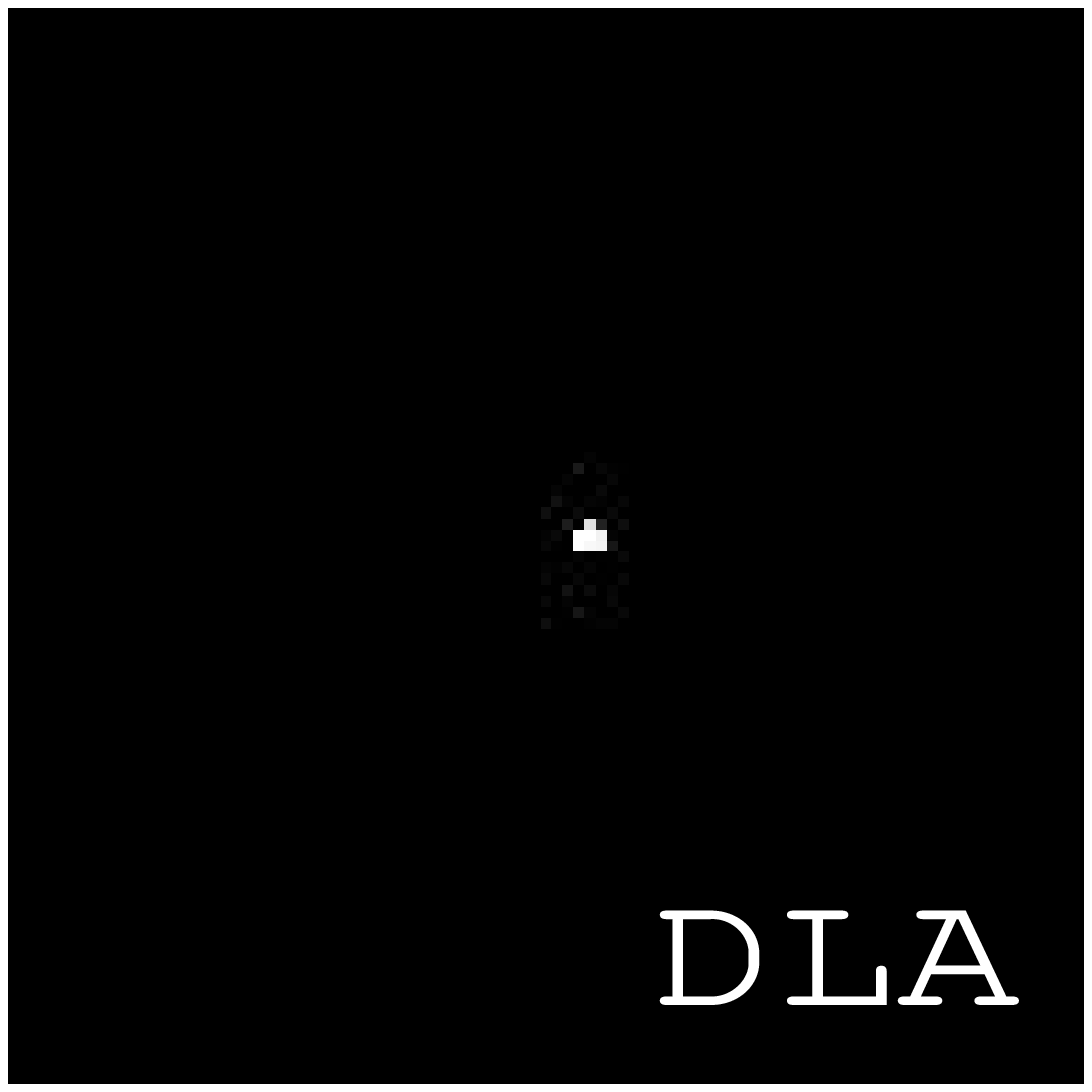}}%
\hspace{0.13cm}
\resizebox{3.8cm}{!}{\includegraphics{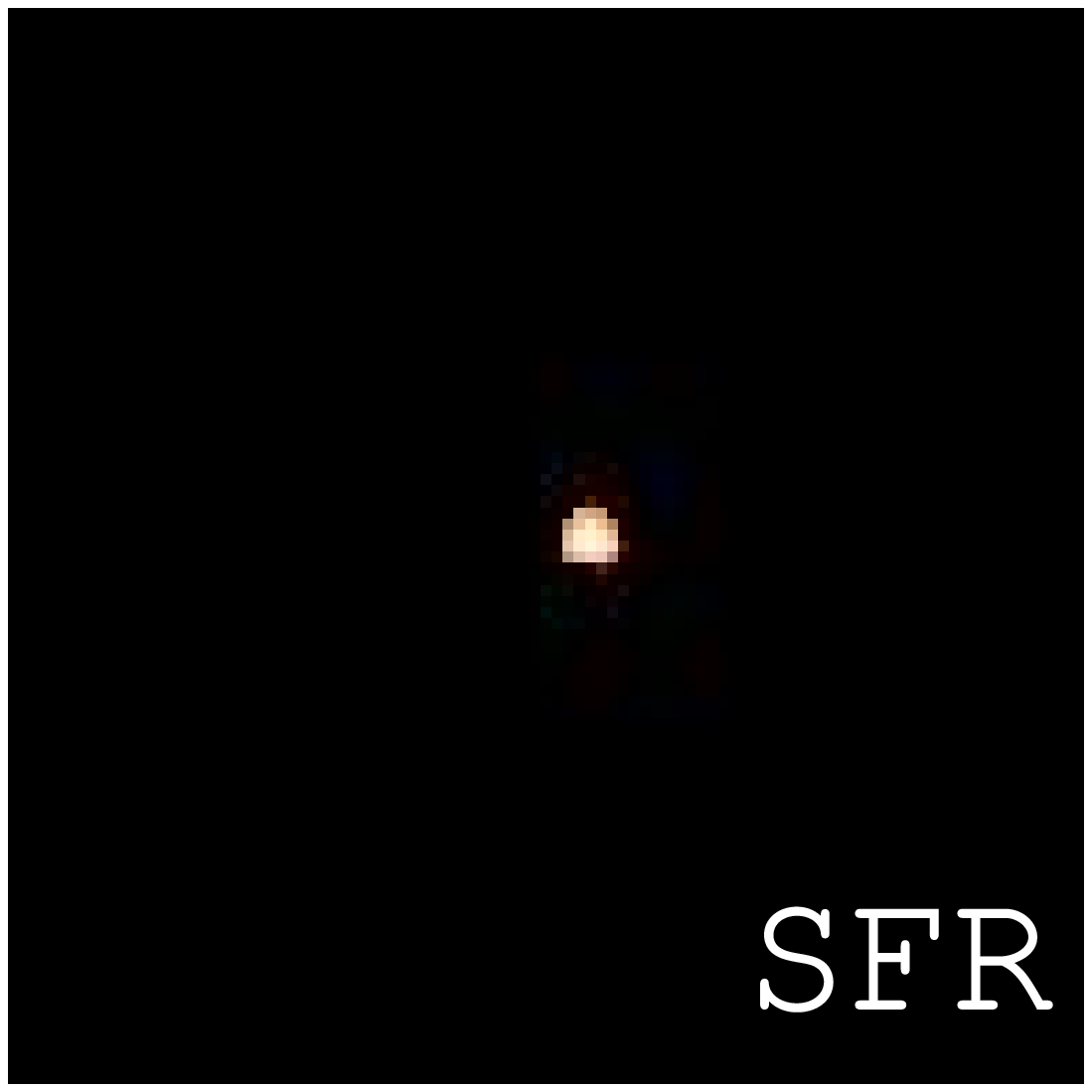}} %
\hspace{0.05cm}
\resizebox{3.8cm}{!}{\includegraphics{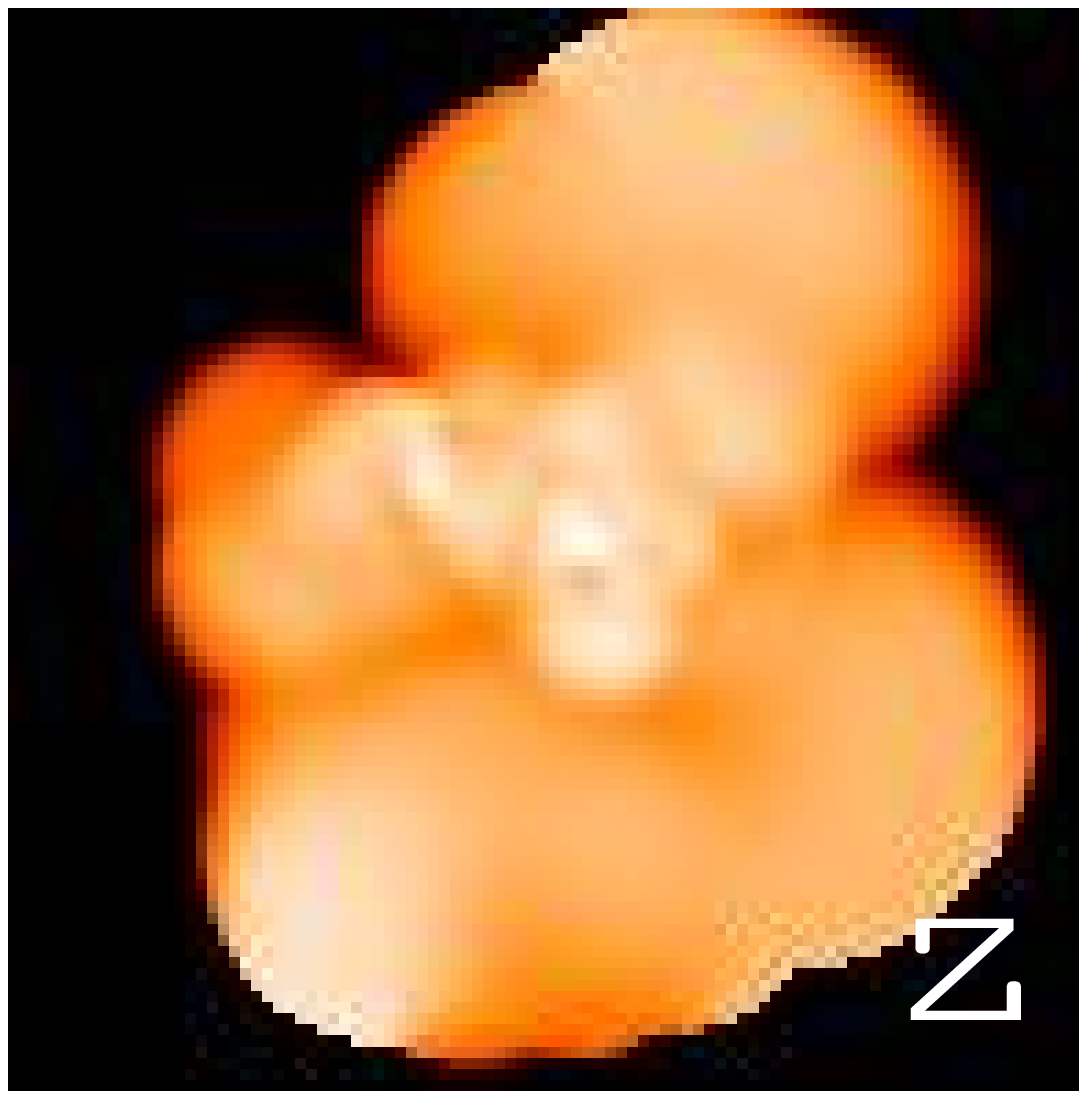}}\\%
\caption{Projected spatial distribution of various quantities for a
halo of mass $\Mhalo=2.4\times 10^{9}\hinv\Msun$ at $z=3$ in the
`Q5'-run.  The size of each panel is comoving $\pm 57 \hikpc$ from 
the centre of the halo.
The graph on the top left shows the probability
distribution function of lines-of-sight (${\rm d}n/{\rm d}\log\NHI$)
for this halo and the one shown in Figure~\ref{dla-pic2.eps} as a
function of $\log\NHI$. One can see that the majority of the orange
region is occupied by lines-of-sight with $11<\log\NHI<16$, which
could be observed in the Ly-$\alpha$ forest. The other panels are ordered
in the same way as in Figure~\ref{dla-pic2.eps}.
\label{dla-pic3.eps}}
\end{center}
\end{figure*}

In Figures~\ref{dla-pic.eps}, \ref{dla-pic2.eps}, and
\ref{dla-pic3.eps}, we show the spatial distribution of the following
quantities for 3 representative haloes in our highest resolution
simulation at $z=3$ (`Q5'-run): neutral hydrogen column density
$\NHI$, DLAs (simply setting a threshold of $\log\NHI>20.3$ to the
total $\NHI$), projected stellar mass surface density, projected star
formation rate surface density, projected metal mass surface density,
and projected metallicity.  The size and the mass of each halo is given in
the caption.  The pixel size of the picture is taken to be the
gravitational softening length of the simulation, which is
$1.23\,\hikpc$ for the `Q5'-run.  We defer a discussion of
quantitative results to the following sections and here simply
summarise notable features of each figure.

Figure~\ref{dla-pic.eps}: This halo is the most massive one at $z=3$
($\Mhalo=1.7\times 10^{12}\hinv\Msun$), which presumably evolves into
a group of galaxies at lower redshift. Note that this is a very rare
halo at $z=3$. With our limited boxsize of $\Lbox=10\himpc$, only a
few such massive haloes (which have roughly the space density of
Lyman-break galaxies) can exist in our simulation box.  The neutral
hydrogen gas distribution in this halo is very extended, and the DLAs
are widely distributed within the halo, but with a trend of higher
concentration towards the centre.  Obviously, the DLA cross-section of
this halo is very high compared to lower mass haloes (see Figure~2 of
\citeauthor{Nag03}, \citeyear{Nag03}).  The inset in the top left
panel shows the probability distribution function of sight-lines
(${\rm d}n/{\rm d}\log\NHI$) for this halo as a function of
$\log\NHI$. One can see that the majority of the red region is
occupied by sight-lines with $11<\log\NHI<15$, which could be observed
in the Ly-$\alpha$ forest.  As the stellar mass density picture shows, a
few high concentrations of stellar mass at the centre would probably
correspond to LBGs at $z=3$ (a detailed analyses of LBGs in our
simulation and their relation to DLAs will be studied elsewhere). The
projected SFR surface density has a similar distribution as the
stellar mass, but they do not necessarily coincide with each other.
The metallicity distribution is still more complex.  While metals are
widely distributed in general, they are also clearly concentrated in
the region of high SFR.  A clear feature is the high metallicity
region (white) at the centre of the halo, where $\NHI$ is also very
high.  An interesting filamentary structure with low metallicity can
be seen in the upper left side of the halo, and there is also a
corresponding feature in the $\NHI$ panel. In this region, there is a
lot of neutral hydrogen, but the gas has not yet been significantly
polluted by star formation, as evidenced by the low stellar mass
density in the same region, explaining the low metallicity.

Figure~\ref{dla-pic2.eps}: This halo of mass $\Mhalo=2.6\times
10^{10}\hinv\Msun$ belongs to a much more abundant class of objects
than the one shown in Figure~\ref{dla-pic.eps}.  As can be seen in the
panel with the stellar mass density (upper middle column), there are 5
or 6 galaxies in this halo, and the left-most clump seems to be going
through a merger. The distribution of DLAs, SFR, and metal
mass all correspond to the five stellar density peaks very
well. Again, the metallicity is high in these galaxies.

Figure~\ref{dla-pic3.eps}: This halo of mass $\Mhalo=2.4\times
10^{9}\hinv\Msun$ is even less massive than the one shown in
Figure~\ref{dla-pic2.eps}, by an order of magnitude. Two peaks
can be seen in the stellar mass surface density distribution, but only
one of them (right one) is a DLA galaxy, and only this galaxy is
forming stars. The one on the left presumably converted all of its
high density cold neutral gas into stars, and therefore no DLA and no star
formation activity can be seen. The metal mass is concentrated in two
galaxies, and the metallicity is high in the centre of the halo.  The
panel on the very left shows the probability distribution function of
sight-lines (${\rm d}n/{\rm d}\log\NHI$) through this halo as a function of
$\log\NHI$, and compares it to the one shown in
Figure~\ref{dla-pic2.eps}. One can see that the majority of the orange
region is occupied by sight-lines with $11<\log\NHI<16$, which
corresponds to the column densities of the Ly-$\alpha$ forest.


\section{\HI column density vs. halo mass}
\label{section:column_mass}

\begin{figure*}
\epsfig{file=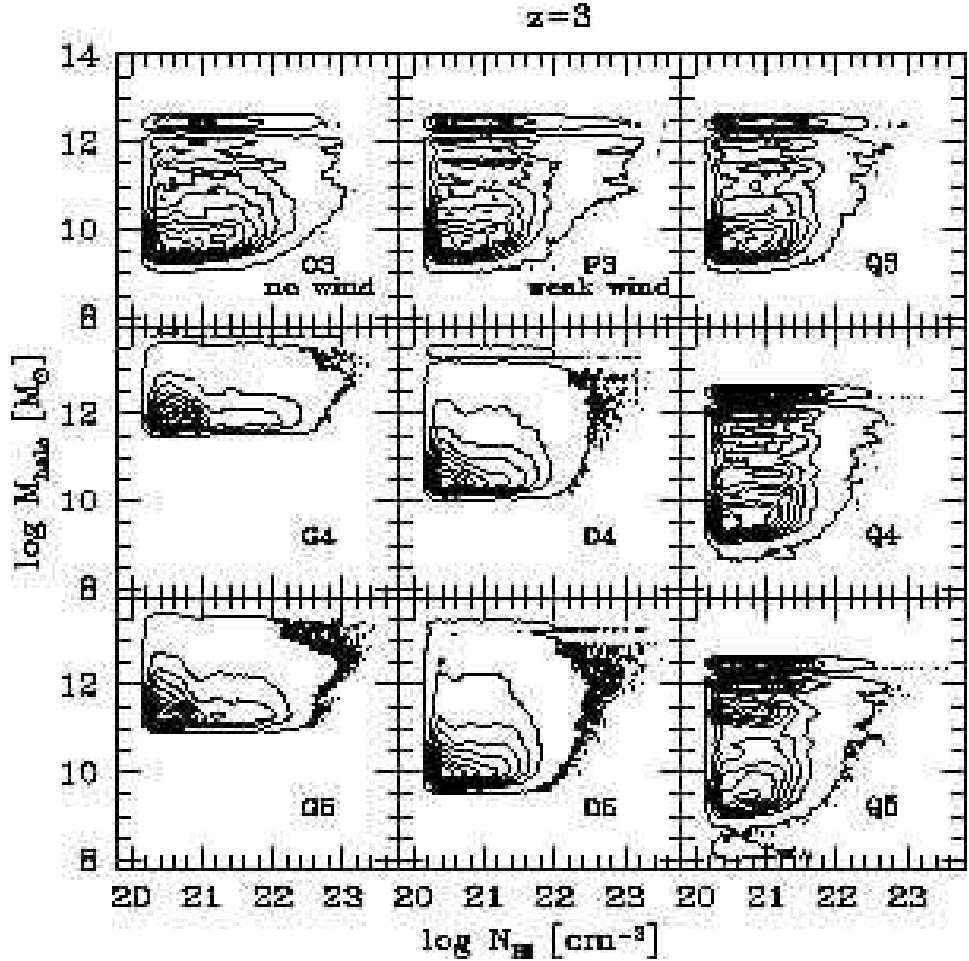,height=5in,width=5in, angle=0} 
\caption{\HI column density vs. dark matter halo mass at $z=3$.  Each
point in the figure represents one line-of-sight. Contours are equally
spaced on a logarithmic scale and the lowest contour contains 95\% of
all the points.  The highest resolution run in this figure is Q5
(bottom left panel).
\label{columnmass_z3.eps}}
\end{figure*}

In Figure~\ref{columnmass_z3.eps}, we show the \HI column density of
DLA sight-lines against the dark matter halo mass in which DLAs
reside.  The calculation of the \HI column density is carried out
exactly the same way as we did in \citet{Nag03}, which we summarise
briefly here.  First, we identify dark matter haloes by applying a
conventional friends-of-friends algorithm to the dark matter particles
in each simulation. After dark matter haloes are identified, we
associate each gas and star particle with their nearest dark matter
particle, including them in the particle list of the corresponding
haloes, when appropriate. Then, for each halo, a uniform grid covering
the entire halo, with grid-size equal to the gravitational
softening length, is placed at the centre-of-mass of the halo.  The
neutral mass of each gas particle is smoothed over a spherical region
of grid-cells, weighted by the SPH kernel. We then project the neutral
gas in the halo onto a plane, and obtain the column density of each
grid-cell in this plane.

We regard each grid-cell on the $xy$-plane as one sight-line, as
we have no power to resolve smaller scales than the gravitational
softening length of each simulation. Each point in the figure 
represents one sight-line. Contours are in equal logarithmic 
scale, and the lowest contour contains 95\% of all the points.

A feature to note here is that a wide range of $\NHI$ values correspond
to massive haloes at $z=3$.  This means that the massive haloes that
host LBGs at $z=3$ also host a wide variety of DLAs.  Note that, since
we identified the dark matter haloes with a simple friends-of-friends
algorithm, subhaloes in massive haloes are not separated from the
parent halo in this figure.  There is also a weak trend that less
massive haloes in general host lower $\NHI$ systems, which is
naturally expected. Low mass haloes with $M_{\rm halo}\leq 10^9 \Msun$
tend to host only one DLA with lower $\NHI$ at the centre of the halo.


\section{Projected stellar mass density and DLAs}
\label{section:starcolumn}

\subsection{Redshift $z=3$}
\label{section:starcolumn_z3}

\begin{figure*}
\epsfig{file=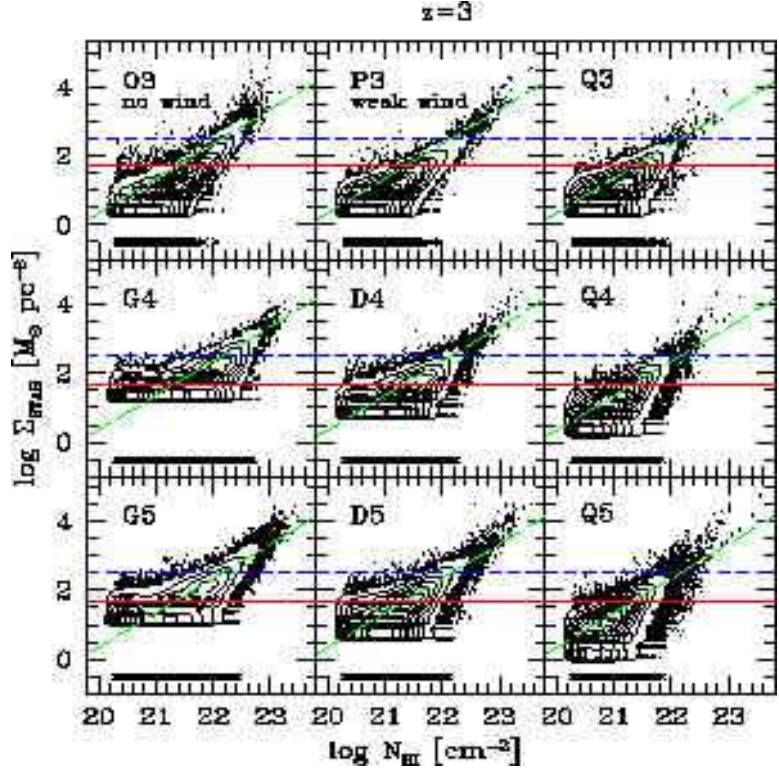,height=4.0in,width=4.0in, angle=0} 
\caption{Projected stellar mass density vs. \HI column density at
$z=3$ (both in proper units, not comoving; same for the rest of the
figures).  Each point in the figure represents one
line-of-sight. Contours are equally spaced on a logarithmic scale and
the lowest contour contains 95\% of all the points.  The highest
resolution run in this figure is Q5 (bottom left panel).  The
long-dashed line indicates the relation of the projected star-to-gas
mass ratio $\Sigstar/\mpro\NHI = 3$.  See text for a discussion of the
other 2 horizontal lines in the figure.
\label{starcolumn_z3.eps}}
\end{figure*}

\begin{figure*}
\epsfig{file=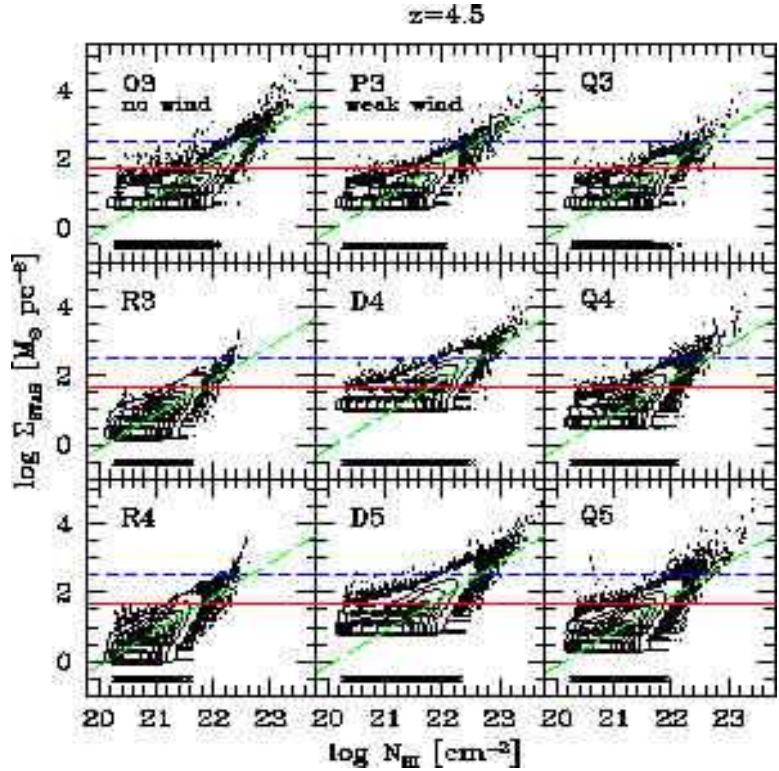,height=4.0in,width=4.0in, angle=0} 
\caption{Projected stellar mass density vs. \HI column density at
$z=4.5$. The long-dashed line indicates the projected star-to-gas mass
ratio $\Sigstar/\mpro\NHI = 1$ (instead of 3 in
Figure~\ref{starcolumn_z3.eps}). The other two horizontal lines are
the same as in Figure~\ref{starcolumn_z3.eps}.
\label{starcolumn_z4.5.eps}}
\end{figure*}

\begin{figure}
\epsfig{file=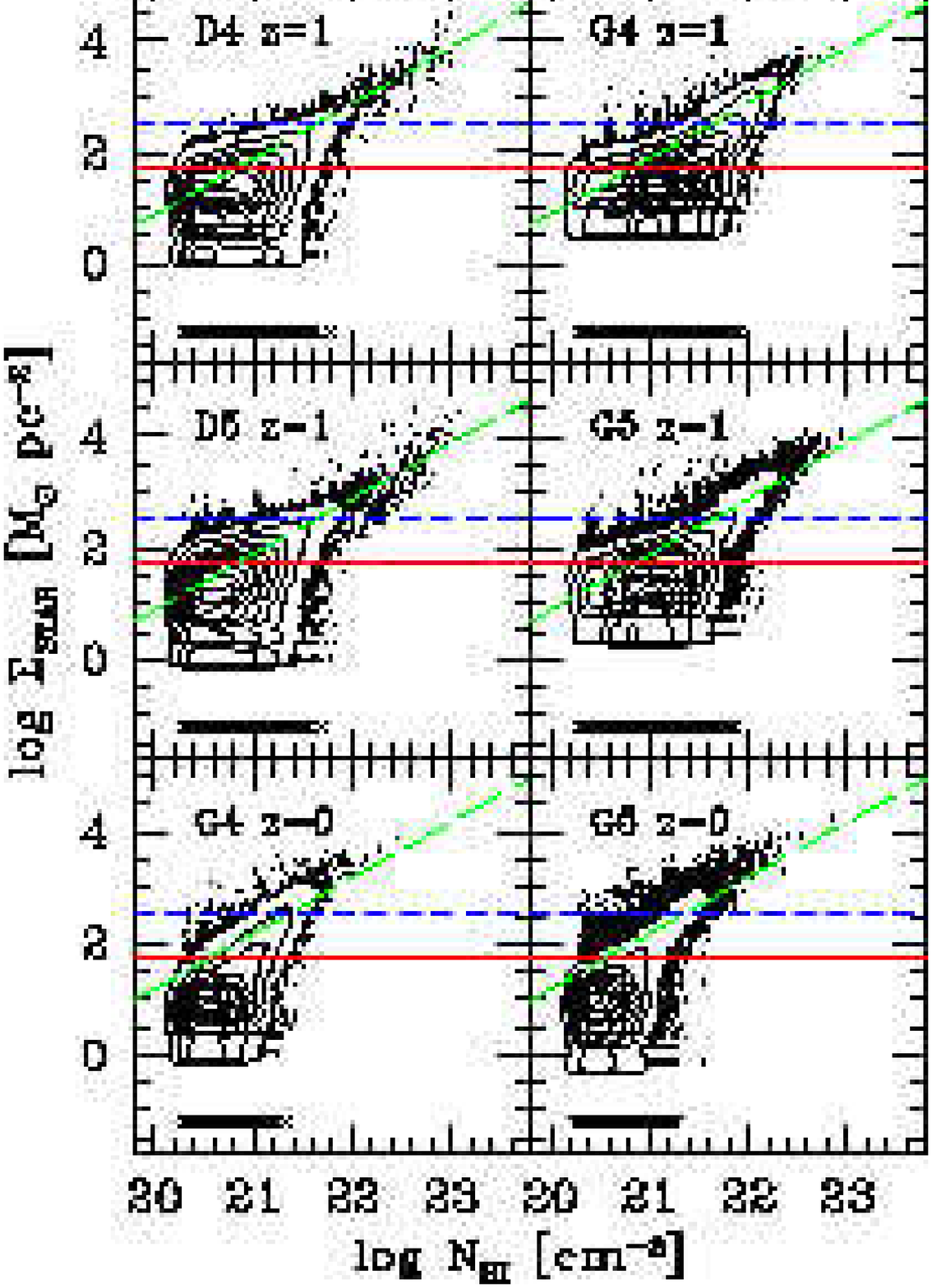,height=4.5in,width=3.2in, angle=0} 
\caption{Projected stellar mass density vs. \HI column density at
$z=1$ and $z=0$. The long-dashed line indicates the projected
star-to-gas mass ratio $\Sigstar/\mpro\NHI = 10$ at $z=1$ (top 4
panels), and 20 at $z=0$ (bottom 2 panels). The other two horizontal
lines are the same as in Figure~\ref{starcolumn_z3.eps}.
\label{starcolumn_lowz.eps}}
\end{figure}

In Figure~\ref{starcolumn_z3.eps}, we show the distribution of
projected stellar mass density (in proper units) of DLAs as a function
of \HI column density.  Each point in the figure represents one
line-of-sight. Contours are in equal logarithmic scale, and the lowest
contour contains 95\% of all the points.  Sight-lines with no stars
are shown as crosses at the bottom of each panel.  As a reference, 
the solid horizontal line indicates an estimate of surface mass density 
of stars in the Milky Way, $\Sigstar = 50~\Msun \pc^{-2}$, although 
this number is highly uncertain because of the difficulty in constraining
the contribution of dark matter in the disc \citep{Bah84, Kui89, Kui91, 
Bah92, Flynn94, Binney98}. Most recent studies using the Hipparcos 
satellite data suggest that there is no compelling evidence for 
significant amounts of dark matter in the disc \citep{Cre98, Hol00}. 
The short-dashed horizontal line indicates a reference value for
$\Sigstar$ if stars of total mass $10^{11}\Msun$ are distributed in
a circle of radius 10 \kpc. As a reference, we show the
projected star-to-gas mass ratio of $\Sigstar/\mpro\NHI = 3$ with the
long-dashed line, where $\mpro$ is the proton mass.  The points are
distributed on both sides of this line at $20<\log\NHI<22.5$, but they
preferentially lie above this line at $\log\NHI>22.5$ in all of the
panels, suggesting efficient star formation at these high column
densities.  Stripes seen at low values of $\Sigstar$ show the
discreteness due to the limited stellar mass resolution of stars in
the simulation.

Comparing the top 3 panels of Figure~\ref{starcolumn_z3.eps}, `O3' (no
wind), `P3' (weak wind), and `Q3' (strong wind), gives us an idea
about the effect of the feedback by galactic winds.  The runs with no
winds (O3) and weak winds (P3) contain more sight-lines with high
$\Sigstar$ when compared to `Q3' (strong wind run, upper right
corner of the panel). This is because the gas in `O3' can cool and
form stars continuously without being blown out by galactic winds.

As we increase the numerical resolution from Q3, Q4, and to Q5 (all
strong wind runs), the number of sight-lines with $\log\NHI>22$
increases, because higher resolution runs can resolve higher gas
density.  Otherwise, the distribution of points in Q3, Q4, and Q5
looks similar.  Comparison between D4 \& D5, and G4 \& G5 also shows a
similar trend.

As the boxsizes become larger from the Q-Series (runs with $10\himpc$
box) to the D- and G-series, the bottom end of the contour rises due
to degrading mass resolution. A larger simulation box (G5) contains
more high-mass haloes than a smaller box (Q5); therefore G5 contains
more sight-lines with high $\Sigstar$ values.


\subsection{Redshift $z=4.5$}

In Figure~\ref{starcolumn_z4.5.eps}, we show the same quantities as in
Figure~\ref{starcolumn_z3.eps} but at redshift $z=4.5$.  Horizontal
solid and short-dashed lines are the same as in
Figure~\ref{starcolumn_z3.eps}. The long-dashed line here shows the
projected star-to-gas ratio of $\Sigstar/\mpro\NHI = 1$.  The
star-to-gas mass ratio at $z=4.5$ is lower than that at $z=3$ by about
a factor of a few.

The highest resolution run in this figure is R4 (bottom right panel),
but the boxsize of this run is small ($3.375\himpc$). A more reliable
distribution can be observed in the Q5 run.


\subsection{Lower redshift}
\label{section:star_lowz}

In Figure~\ref{starcolumn_lowz.eps}, we show the same quantities
as in Figure~\ref{starcolumn_z3.eps} but at redshifts $z=1$ and $z=0$. 
The two horizontal lines are the same as those in 
Figure~\ref{starcolumn_z3.eps}. The long-dashed line 
indicates the projected star-to-gas mass ratio $\Sigstar/\mpro\NHI = 10$ 
at $z=1$ (top 4 panels), and 20 at $z=0$ (bottom 2 panels).

The observational estimate of the local star-to-gas mass ratio 
in the Milky Way is $\Sigstar/\mpro\NHI \approx 10$. This is obtained
by dividing the estimate of $\Sigstar \approx 50~\Msun \pc^{-2}$ 
(assuming that there is no significant contribution of dark matter 
in the disc) \citep{Bah84, Kui89, Kui91, Bah92, Flynn94, Binney98, 
Cre98, Hol00} by the contribution from 
the interstellar medium, $\Sigism \approx 5~\Msun \pc^{-2}$ \citep{Dame}.
Given the uncertainties in these numbers, the values we find in the 
simulation are consistent with the observed range.

An interesting evolutionary feature in the distribution can be
observed as a function of redshift. At $z=3$, the distribution is wide
in the horizontal direction, but as the redshift decreases, the
distribution becomes narrower and elongated in the vertical direction, and
the points move to higher $\Sigstar$ and lower $\NHI$ values as the
neutral gas is converted into stars inside the galaxies.


\section{Star formation rate of DLA galaxies}
\label{section:sfrcolumn}

\begin{figure*}
\epsfig{file=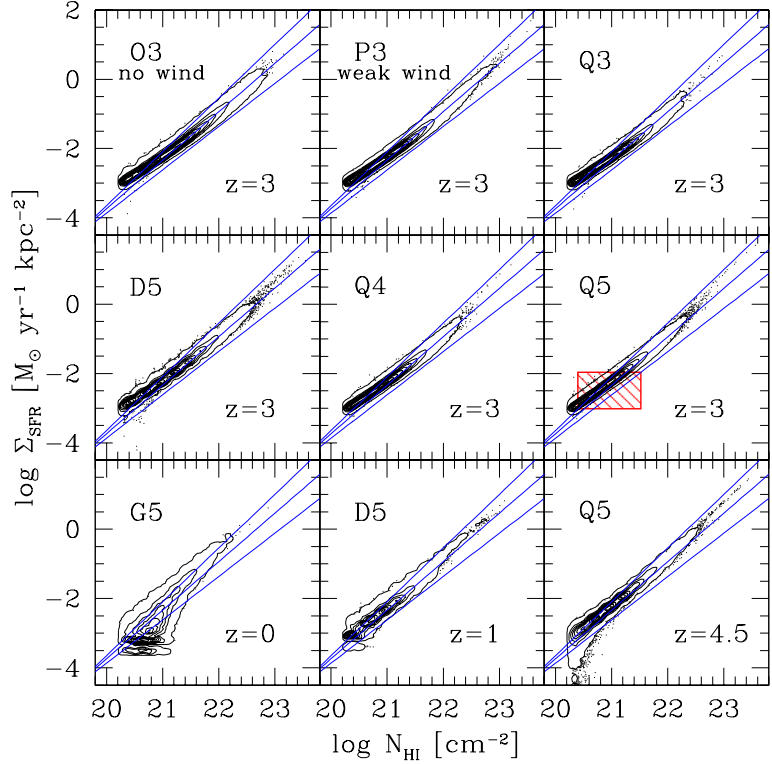,height=4in,width=4in, angle=0} 
\caption{Projected star formation rate density vs. \HI column density
at $z=0$, 1, 3, \& 4.5. Each point in the figure represents one line-of-sight.
Contours are equally spaced on a logarithmic scale and the lowest
contour contains 95\% of all the points.  The shaded region in the
Q5 panel roughly indicates the region of recent observational data 
by Wolfe et al.~(2003b).
\label{sfr_nh_all.eps}}
\end{figure*}

In Figure~\ref{sfr_nh_all.eps}, we show the projected star formation
rate surface density $\Sigsfr$ (in proper units of
$\Msun\yr^{-1}\kpc^{-2}$) as a function of \HI column density.  The
SFR of each gas particle is smoothed over a spherical region of
grid-cells, weighted by the SPH kernel.  This ensures that rate of
star formation is conserved when added up along the $z$-direction to
obtain the projected SFR density.

Each point in the figure represents one line-of-sight.  Contours are
equally spaced on a logarithmic scale, and the lowest contour contains
95\% of all the points.  The distribution was truncated at
$\log\NHI=20$, so the cutoff of the contours at $\log\NHI\sim 20$ is
artificial.  The three solid lines in each panel show the allowed
region by the following empirical relation between the projected SFR
density and the \HI column density, known as the Kennicutt law
\citep{Ken98}: \beq
\label{eq:kennicutt}
\Sigsfr = (2.5\pm0.7)\times 10^{-4} \rp{\frac{\NHI}{1.25\times 10^{20}\cm^{-2}}}^{1.4\pm0.15} \\
\Msun\yr^{-1}\kpc^{-2}.
\eeq

The projected SFR of DLAs in the simulation agrees well with the
Kennicutt law. It may at first be surprising that the simulated result
at $z=3$ follows the Kennicutt law so well, which empirically is only
established for $z=0$. We will discuss this point further in
Section~\ref{section:discussion}.  There is a slight trend in all
panels for the distribution to deviate to higher values of $\Sigsfr$
at $20<\log\NHI<21$ relative to the Kennicutt law, but the reason for
this deviation is not clear.

The shaded area in the panel of Q5 in Figure~\ref{sfr_nh_all.eps} 
roughly indicates the region of recent observational data by \citet{Wol03b}. 
Their data points are close to the simulated result and roughly falls on 
the Kennicutt law, but the individual measurements do not with some scatter.
Wolfe et al. (2003b) argue that this scatter is not surprising 
because their measurement of projected SFR is an average over a scale 
of a couple of $\kpc$ over which $\NHI$ is expected to fluctuate.

As the wind strength increases from O3 to Q3, the number of high
$\NHI$ points decreases for the same reason described in
Section~\ref{section:starcolumn_z3}, but otherwise the distribution
remains very similar for different wind models.  This is because even
though the \HI content of galaxies can be changed by a galactic wind,
the star formation recipe in the code gives the same SFR for a given
\HI content. Varying the wind strength therefore only moves the points
along the Kennicutt law.

Increasing the numerical resolution from Q3, Q4, and then to Q5,
slightly increases the number of sight-lines at high $\NHI$ values,
but the distribution is basically identical. In the G-series, one sees
a sprinkle of points that fall away from the Kennicutt law at around
$20<\log\NHI<21$, which is perhaps due to a lack of resolution in
these large boxsize runs.

\begin{figure*}
\epsfig{file=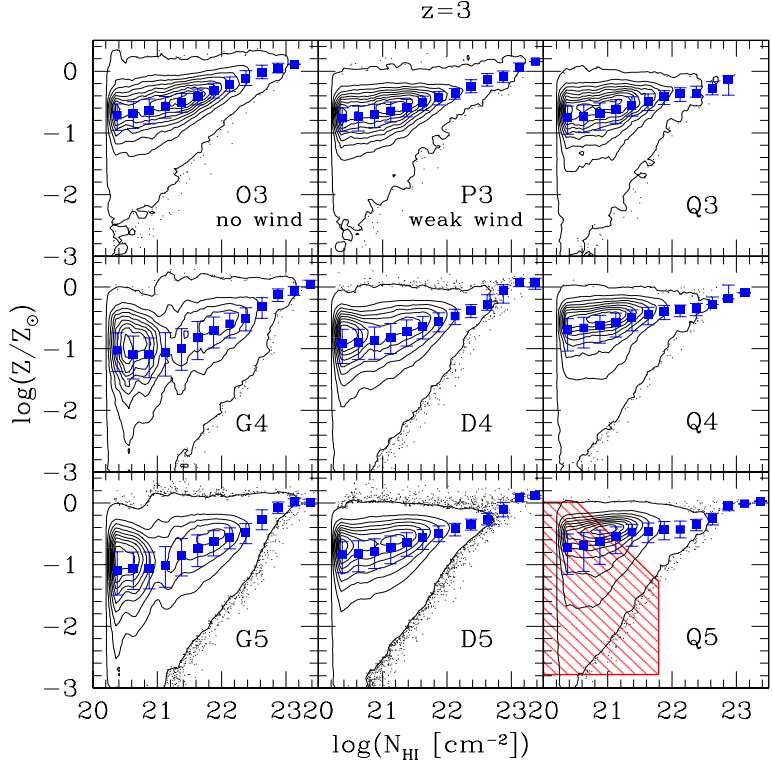,height=4in,width=4in, angle=0} 
\caption{Gas metallicity vs. \HI column density at $z=3$.  Each point
in the figure represents one line-of-sight. Contours are equally
spaced on a logarithmic scale.  Current observational data points fall
into the shaded region that is drawn in the bottom left panel (Q5).
\label{metalcolumn_z3.eps}}
\end{figure*}

At $z=4.5$ (only Q5 is shown), the distribution of points seems to be 
tighter than that at $z=3$.  An interesting feature seen at $z=4.5$ is 
the significant number of
points falling away from the Kennicutt law at $\log\NHI\sim 20.5$.
This is indicating that there are many more DLA sight-lines with
inefficient star formation at $z=4.5$ compared to $z=3$, which is
presumably related to the sudden decline in the DLA cross-section at
around $\Mhalo=10^{8.3}\hinv\Msun$, as found in \citet{Nag03}. At
$z=4.5$, many low-mass haloes still have not accumulated enough gas
for them to radiatively cool, and the neutral gas content and the SFR
is significantly lower than in those at $z=3$.

At low redshift ($z=0$ \& 1), the distribution starts to shrink towards lower
$\Sigsfr$ and lower $\NHI$ values. This can be largely understood in
terms of the global evolution of the star formation efficiency in
haloes, and its variation with haloes mass. As the redshift decreases, a
larger fraction of the total star formation rate in the universe is
contributed by ever more massive haloes \citep{SH03b}.  In addition, at
lower redshifts, haloes are less dense, and their gas has been heated by
prior star formation, such that cooling becomes comparatively
inefficient \citep{Blanton99}, leading to a reduction in the star
formation rates.


\section{Gas Metallicity of DLAs}
\label{section:metalcolumn}

\subsection{Redshift z=3}

In Figure~\ref{metalcolumn_z3.eps}, we show the projected gas
metallicity as a function of \HI column density at $z=3$.  The
projected metallicity was calculated as follows: First, the metal mass
and gas mass of each gas particle was smoothed over a spherical region
of grid-cells, weighted by the SPH kernel. We then added up gas and
metal mass independently along the $z$-direction, and finally divided
the sum of the metal mass by the sum of the gas mass for each
sight-line. We have also tried the same calculation by weighting with
the neutral hydrogen mass instead of using the total gas mass, but the
result was almost identical.  We also confirmed that the result is
robust even when we select out only the gas particles in high density
clumps, therefore the results are not affected by the diffuse gas in
the halo when projected along the $z$-direction. 

Each point in the figure represents one sight-line, and the contours
are equally spaced on a logarithmic scale. The solid square symbols
give the median value in each $\log\NHI$ bin, with error bars
indicating the quartiles on both sides. In all panels, the median
metallicity increases as $\NHI$ increases, reaching solar metallicity
at $\log\NHI\sim 23$.  This trend is expected because star formation
is more vigorous in high $\NHI$ systems, as we saw in
Section~\ref{section:sfrcolumn}.

In the `O3' run (no wind), the contours extend to higher metallicity
than in the `Q3' run (strong wind). This is because in O3 star
formation is not suppressed by a galactic wind.  Therefore, the gas is
more enriched with metals due to a higher average star formation
rate. The values of the median metallicity at intermediate column
densities ($20<\log\NHI<22$) are similar for all runs in the Q-series
(O3, P3, Q3, Q4, and Q5).

Increasing the resolution from Q3, to Q4, and finally to Q5, increases
the number of points with high metallicity at high $\NHI$ values, but
otherwise the overall distribution is very similar.

It is clear that simulations with larger boxsizes, such as those of
the D- and G-series, underestimate the metallicity for systems with
$\log\NHI<22$ at $z=3$ due to lack of resolution. This is consistent
with the finding that in the runs of the D- and G-series the \HI
column density distribution function at $z=3$ is not reliably
estimated, as was shown in \citet{Nag03}.

An important result is that the median metallicity we find for DLAs in
our simulations is much higher than the observed value. Note that
observers typically find DLAs with $20<\log\NHI<22$ to have a
metallicity of $Z/\Zsun \sim 1/30$ \citep[e.g.][]{Boisse98, Pettini99,
Pro00}, as indicated by the shaded region in the panel of `Q5' (bottom
left).  We will discuss the implications of this discrepancy further
in Section~\ref{section:discussion}.

\begin{figure}
\epsfig{file=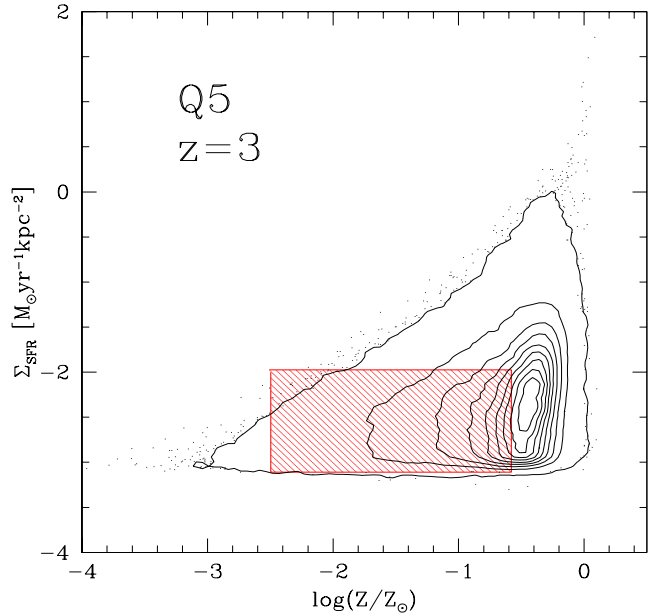, height=3.2in, width=3.35in, angle=0} 
\caption{Projected SFR vs. gas metallicity at $z=3$ for the `Q5'-run.
Each point in the figure represents one line-of-sight. Contours are
equally spaced on a logarithmic scale. The shaded area indicates the
region of observed data points by Wolfe et al.~(2003b).
\label{sfr_metal_z3.eps}}
\end{figure}

In Figure~\ref{sfr_metal_z3.eps}, we show the distribution of SFR as a
function of metallicity at $z=3$ for the `Q5'-run.  The recent
observational data points by \citet{Wol03b} fall into the shaded
area.  The shape of the simulated distribution is easy to
understand: Because $\Sigsfr$ is tightly correlated with $\NHI$, as
we saw in Figure~\ref{sfr_nh_all.eps} (the Kennicutt law), the distribution
seen in Figure~\ref{sfr_metal_z3.eps} is simply a reflection of
Figure~\ref{metalcolumn_z3.eps} around a diagonal line.

\subsection{Redshift $z=4.5$}

\begin{figure*}
\epsfig{file=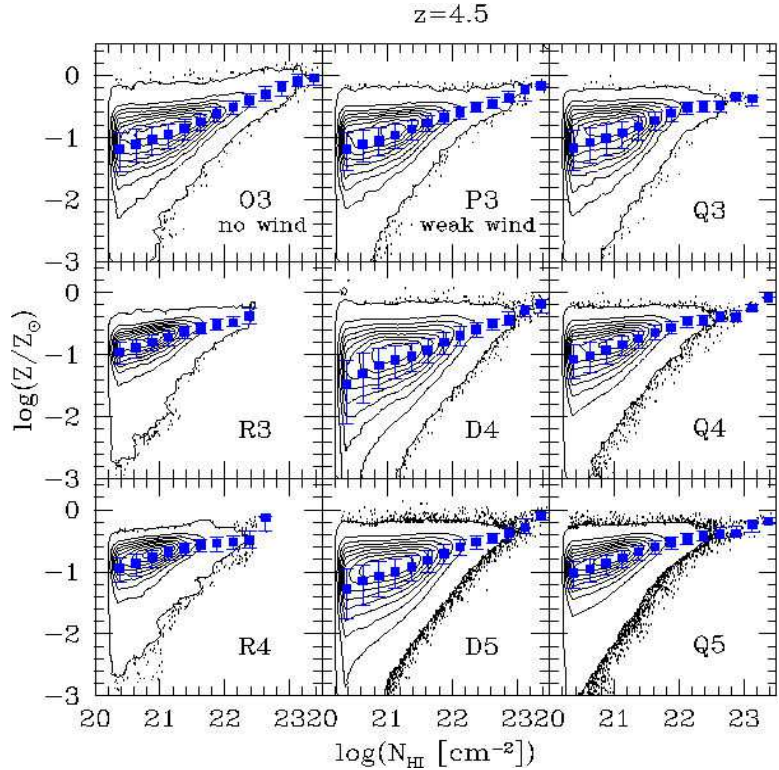, height=4in,width=4in, angle=0} 
\caption{Gas metallicity vs. \HI column density at $z=4.5$.  Each
point in the figure represents one line-of-sight. Contours are equally
spaced on a logarithmic scale.
\label{metalcolumn_z4.5.eps}}
\end{figure*}

In Figure~\ref{metalcolumn_z4.5.eps}, we show the same quantities as
in Figure~\ref{metalcolumn_z3.eps}, but for $z=4.5$.  A noteworthy
feature is that the median metallicity at $z=4.5$ is lower than that
at $z=3$ by a factor of $2-3$ at column densities $\log\NHI<22$. This
is consistent with the fact that the star-to-gas mass ratio at $z=4.5$
is lower than that at $z=3$ by a similar factor.  The trends observed
for increasing numerical resolution and wind strength are similar to
those at $z=3$.


\subsection{Lower redshift}

\begin{figure}
\epsfig{file=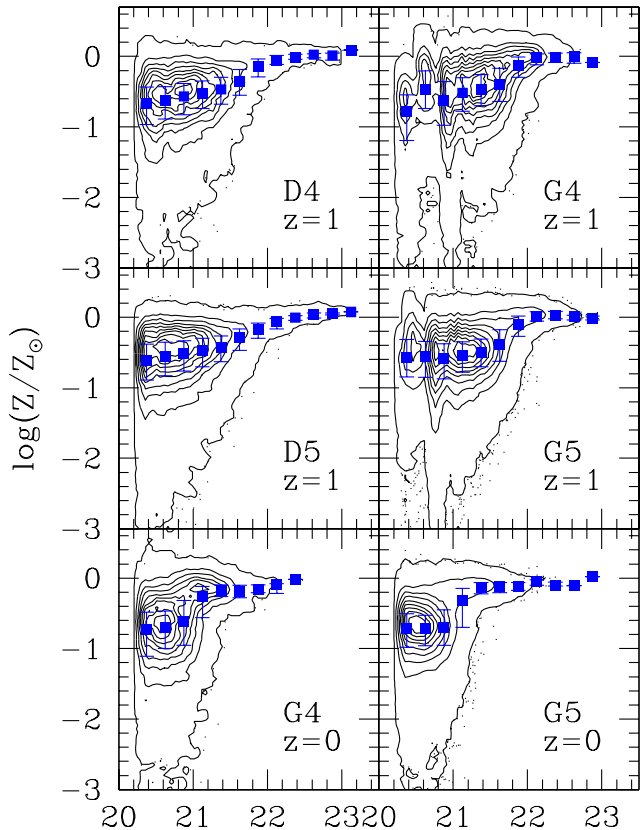, height=4.3in, width=3.3in, angle=0} 
\caption{Gas metallicity vs. \HI column density at $z=0$ and 1.  Each
point in the figure represents one line-of-sight. Contours are equally
spaced on a logarithmic scale.
\label{metalcolumn_lowz.eps}}
\end{figure}

In Figure~\ref{metalcolumn_lowz.eps}, we again analyse the same
quantities as in Figure~\ref{metalcolumn_z3.eps}, but at redshifts 
$z=1$ and $z=0$.  The most reliable of our runs at $z=1$ is `D5', which
shows a median metallicity that is higher by a factor of $\sim 2$
compared to that at $z=3$.  A factor of 3 increase is observed for
`G5'.

We note that the evolution in the median metallicity seen from $z=1$
to $z=0$ is quite uncertain.  In fact, for the `G5' run, the 3 median
points at $20<\log\NHI<21$ go down by a factor of 2 from $z=1$ to
$z=0$. This is because in a simulation with a large boxsize, such as
`G5', the mass resolution is low and many haloes are just forming at
low redshift. Therefore, the number of points at $20<\log\NHI<21$ with
low metallicity is still increasing from $z=1$ to $z=0$, as evidenced
by the broadening of the contours towards lower metallicity. This is
consistent with the fact that the value of $\OHI$ in `G5' slightly
increases from $z=1$ to $z=0$, as was shown in Figure~1 of
\citet{Nag03}.

\section{Mean Metallicity Evolution}
\label{section:mean-metal}

\begin{figure}
\epsfig{file=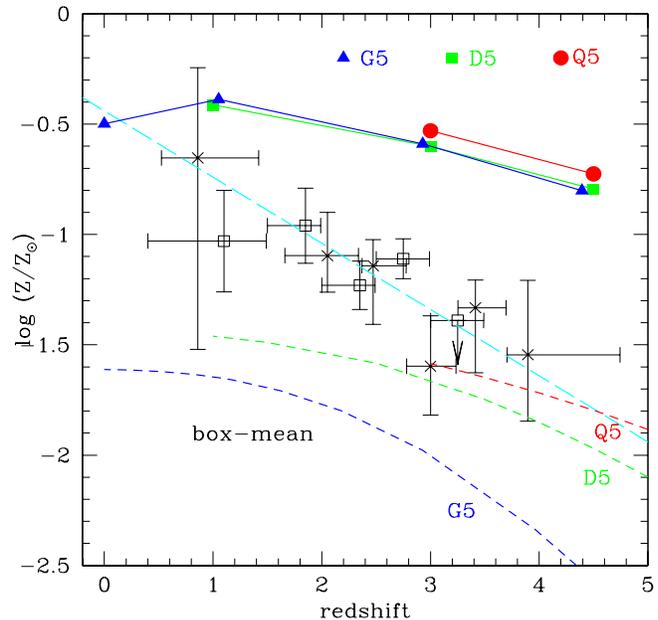, height=3.2in, width=3.35in, angle=0} 
\caption{Evolution of the mean gas metallicity of DLAs as a function
of redshift. The symbols connected by solid lines give the direct
simulation results for the Q5, D5, and G5-runs.  The short-dashed 
lines at the bottom of the figure are the mean metallicities of the 
entire simulation boxes. Data points with error bars show observations 
by \citet{Pettini99} (open squares) and \citet{Pro03} (crosses).
The long-dashed line is the best-fit line obtained by performing a
least-square fit to the mean points of \citet{Pro03}, which 
exhibits a mild evolution with a slope of $-0.30$ as a 
function of redshift.
\label{metal_bulk.eps}}
\end{figure}

In Figure~\ref{metal_bulk.eps}, we compare observations of the
evolution of the $\NHI$-weighted mean metallicity of DLAs 
with results measured from our simulations. Here, by ``$\NHI$-weighted'',  
we mean summing up all the metal mass in DLA sight-lines and 
dividing it by the total gas mass in all DLA sight-lines.
The symbols connected by the solid lines are the
direct results of our simulations for DLAs in the Q5, D5, and G5-runs.
We have also tried to do the calculation in a different manner; 
simply taking the mean of the metallicities of all lines-of-sight 
in the simulations; i.e.~$\avg{Z/\Zsun} = \Sigma (Z/\Zsun)/N$, 
where $N$ is the total number of lines-of-sight. 
This method gives a mean metallicity which is lower by
$\sim 0.1$ dex than the one obtained by the former method. This is
perhaps because of the wide range of values that DLA metallicities can
take, as seen in Figure~\ref{metalcolumn_z3.eps}.  Our conclusion
regarding the DLA metallicity does not change depending on which
calculation method is used.

Column density-weighted mean observational data points by \citet{Pettini99} 
(open squares) and \citet{Pro03} (crosses) are also shown. 
Note that the error bars of \citet{Pettini99} points are 1-$\sigma$, 
and those of \citet{Pro03} are 95\% confidence level uncertainty, both 
determined from a bootstrap error analysis.
The long-dashed line is the best-fit line obtained by performing a
least-square fit to the mean points of \citet{Pro03}, which 
exhibits a mild evolution with a slope of $-0.30$ as a 
function of redshift [$\log (Z/\Zsun) = -0.30 z - 0.44$].
Note that \citet{Pro03} performed a least-square fit to the central 
values of the error bars, so they report a slope of -0.26 which is 
slightly different from our value, but the two agrees with each other 
within 1-$\sigma$.

The mean metallicity of DLAs in our simulations is again higher than
current observations by about a factor of $3-5$.  On the other hand,
the mean metallicity of the entire simulation box (i.e. total metal mass
divided by the total gas mass) is lower than the
observed DLA metallicity. This makes sense because the box-mean
includes low metallicity IGM that is not polluted as strongly as
galaxies themselves. This simple consideration also illustrates that
the metallicity values of DLAs can be quite sensitive to details of
how metals are transported and mixed with gas in the haloes of
galaxies and with gas in the IGM. In fact, we find that our simulations
predict the correct metallicity of Lyman-$\alpha$ forest when the winds 
are included (Springel et al. 2003, in preparation), and that the box-mean
metallicity is biased higher relative to that of the low-density IGM.

Residual differences between the results of Q5, D5, and G5 are due to
numerical resolution. Higher resolution runs (like Q5) can resolve
smaller haloes at high redshift than lower resolution runs (D5, G5).
Therefore, Q5 has more star formation, and consequently more metal
production, at high redshifts than D5 and G5.

Another noteworthy result is that the $rate$ of the evolution of mean
metallicity in our simulations appears to be consistent with the most
recent observed rate by \citet{Pro03}.  In the simulation of
\citet{Cen02}, a similar agreement was obtained.  We will discuss this
more in the following section.


\section{Discussion}
\label{section:discussion}

We have used a series of state-of-the-art hydrodynamic simulations to
study the distribution of star formation rate and metallicity of DLAs.
Our simulations suggest that the median projected star-to-gas mass
ratio of DLAs is about a few at $z=3$, and then evolves to higher
values of $\simeq 10$ at $z=1$ and $\simeq 20$ at $z=0$. However, we
note that our low-redshift results should be regarded with caution
because they are based on simulations with coarser resolution than
those at high-redshift. Nevertheless it is encouraging that the
star-to-gas mass ratio we find at $z=0$ is close to observational
estimates for the Milky Way (see Section~\ref{section:star_lowz}).

The projected SFR density as a function of neutral hydrogen column
density follows the Kennicutt law well at all redshifts.  At first
sight, this may seem non-trivial and perhaps somewhat surprising.
However, the star formation model adopted in the simulations depends
only on local physical quantities, e.g.~the local gas density. The
free parameter of the model was chosen to reproduce the Kennicutt law
in isolated disk galaxies at low redshift, but was kept fixed as a
function of time. We hence implicitly assumed that the Kennicutt law
holds at all redshifts, and our simulation results reflect this
assumption.

A recent study by \citet{Kra03} also suggests that the projected SFR
follows the global `Schmidt' law, and that it does not depend on
redshift. He argues that the global Schmidt-law originates from a
generic power-law distribution of the high-density tail of the
probability distribution function of gas density.  If the star
formation at high-redshift follows similar physical process as in the
Local Universe, it is not unreasonable that the projected SFR 
should follow the Kennicutt law well even at high redshift.

The \HI column density distribution in our simulations extends up to
values as high as $\log\NHI \sim 23$, which is in the range of \HI
column densities seen in the nuclei of starburst galaxies in the Local
Universe. This suggests that the intense star-formation seen in some
of the brightest simulated galaxies at $z=3$ is similar to the
star formation in local starburst galaxies.  These simulated galaxies
should then correspond to the observed Lyman-break galaxies, which are
often estimated to have SFRs of $\sim 50-100\Msun\yr^{-1}$, or even
higher.  The spatial correlation between Lyman-break galaxies and DLAs
is hence of significant interest to elucidate this connection further,
and we will study this issue in future work.

While the star-to-gas mass ratio and the projected SFR density at
$z=3$ in the simulations seems to be plausible, the median metallicity
of DLAs at $z=3$ appears to be too high ($Z/\Zsun \sim 1/3$) compared
to typically observed values of $Z/\Zsun \sim 1/30$ for DLAs
\citep[e.g.][]{Pettini99}. There are a number of possible explanations
for this problem, and we will briefly discuss some of the most
prominent possibilities.

Since the SFRs and metallicities of galaxies in the simulation at
$z=3$ seem plausible, one possible reason for high metallicity in DLAs
is that the feedback by galactic winds is not efficient enough in
blowing out metals from DLAs. Clearly, if the feedback by winds were
stronger, then star formation and hence metal creation in DLAs would
be more strongly suppressed.  In fact, the SFR of systems with
$20<\log\NHI<21$ deviates slightly from the Kennicutt law to higher
values in our simulation. Reducing the SFR of these systems would
bring them down to values consistent with the Kennicutt law.

However, simply making the winds stronger and blowing out more gas
will not necessarily decrease the metallicity of DLAs much, because in
our current simulation model, the wind transports away metals {\em
and} gas at the same time, i.e.~the wind's initial metallicity is
assumed to be equal to that of the gas of the DLA, leaving the ratio
of metal and gas mass in the DLA unchanged.

It is however quite plausible that the wind is {\em metal-loaded}
compared to the gas in the DLA, as is for example suggested by
simulations of SN explosions (e.g. MacLow \& Ferrara 1999,
Bromm et al. 2003). After all, the ejecta 
of SN are heavily enriched and inject large parts of
the energy that is assumed to ultimately drive the outflow. If the
mixing with other DLA-gas is not extremely efficient before the
outflow occurs, it can then be expected that the wind material has
potentially much higher metallicity than the DLA, thereby selectively
removing metals.

A related possibility concerns the metallicities of the cold and
diffuse phases of the DLA. In the present study, we assumed that
metals are always efficiently and rapidly mixed between the gas of the
cold clouds and the ambient medium, such that there is a homogeneous
metal distribution in the DLA (operationally, we used only a single
metallicity variable for each gas particle, reflecting this
assumption).  However, this assumption may not be fully correct.  If
the metals were preferentially kept in the hot phase of the ISM after
they are released by SN, then they would not be observed in the cold
gas that is responsible for the DLAs. Since we did not track the metal
distribution in cold and hot phases separately in the current
simulations, we may then have overestimated the amount of metals in
DLAs by counting those in the hot phase as well as those in the cold
phase. Note that a more detailed tracking of metals in the simulation,
separately for hot and cold phases of the gas, could in principle be
done easily on a technical level. The difficulty however lies in
obtaining a reasonable description of the physics that governs the
exchange of metals between the different phases of the ISM, something
that is presently not attainable from either observation or theory.
     
The viability of the feedback model in the simulations can also be
tested by comparing with observations of the Lyman-$\alpha$ forest,
which is generated by systems of much lower column density than the
DLAs studied in this work. Curiously, an analysis using the current
simulation series (Springel et al. 2003, in preparation), as well as a
study by \citet{Theuns02}, suggest that the spectral features of the
Ly-$\alpha$ forest are not significantly affected by the feedback from
galactic outflows, despite the fact that the wind strength is taken
to be on the `strong side' in these studies, and despite the fact that
a non-negligible fraction of the IGM volume is heated by the
winds. Note that in our simulations with a strong wind model, the \HI
mass density in the entire simulation box is somewhat lower than
suggested by observational estimates \citep{Nag03}, therefore it
appears problematic to increase the wind strength even beyond the
present value.

The high metallicity of DLAs may also be related to the steep
luminosity function of galaxies in our SPH simulations.  A preliminary
analysis using a population synthesis model shows that the luminosity
function still has a very steep slope at the faint end even at low 
redshift, similar to that of the dark matter halo mass function.  
(It is not obvious whether the steep faint end in the simulation 
at $z>1$ is a problem, because it is not well constrained 
observationally at $z>1$ yet.)  This means that the
formation of low-mass galaxies in our simulation was not suppressed
enough, or equivalently, that star formation was too efficient in
low-mass haloes. Stronger winds may help to alleviate this problem,
but it appears unlikely that our present feedback model can solve it
satisfactorily simply by adopting a higher efficiency parameter for
feedback. It is more plausible that additional physical processes need
to be considered in a more faithful way. One simple possibility for
this is related to the UV background field, which is turned on by hand
at $z=6$ in our present simulations to mimic reionisation of the
Universe at a time when the first Gunn-Peterson troughs in spectra to
distant quasars are observed \citep{Becker01}. However, it is possible
(in fact suggested by the WMAP satellite) that the Universe was
reionised at much higher redshift. The associated photoheating may
have then much more efficiently impaired the formation of low-mass
galaxies than in our present simulations \citep[but see][]{Dij}.  
We plan to explore this possibility in future work by adopting 
different treatments of the UV background radiation field.

It is also interesting to compare our results with other hydrodynamic
simulations of DLAs.  \citet{Cen02} have recently studied the
metallicity of DLAs in an Eulerian hydrodynamic simulation of boxsize
$\Lbox=25\himpc$.  The median metallicity as a function of \HI column
density in their simulation is lower than ours by a factor of $\sim
5$, in better agreement with observations (but still higher than
observations by about a factor of a few). This difference between the
two simulations could be due to the difference in the efficiency of
the SN feedback, which in turn originates in the widely different
spatial resolutions of the two simulations.  The Eulerian simulation
of Cen et al. used a fixed grid, where the energy and the metals released
by SN were dumped into the gas of those local cells that showed star
formation activity.  Since the comoving cell size of their simulation
was $32.6 \hikpc$, metals were thus distributed efficiently over this
spatial scale by construction, diluting the metal mass relative to the
gas mass in the cell for a given amount of stars formed.  On the other
hand, in our present SPH simulations, the gravitational softening
length was typically only a few kpc comoving, such that the gas
particles would in fact have to travel over a large number of
resolution elements in order to distribute the injected metals over
similar spatial scales as in the mesh simulation.

Despite the differences in the metallicity values of DLAs measured
in the two simulations, there are also important common features. One
of them is the existence of high $\NHI$ systems with high metallicity,
which are absent in current observations.  The existence of such a
population of DLAs may perhaps be a generic prediction of current CDM
simulations.  \citet{Cen02} invoked a dust obscuration effect in order
to reconcile their result with observations, and argued that their
column density distribution, \HI mass density, and DLA-metallicity
would all become consistent with observations, provided a sizable 
dust extinction is assumed.  However, it is presently not very clear
how strong dust effects really are \citep[e.g.][]{Ellison, Pro02}, 
and the solution could rather lie
in a more adequate treatment of star formation and supernova
feedback, as \citet{Cen02} also warn.  For example, \citet{Schaye}
argues that the conversion of neutral hydrogen atoms into a molecular 
form, which we have not yet implemented in our simulations, 
would introduce a physical limit to the highest $\NHI$
that DLAs can attain. This process would eliminate the highest
$\NHI$ systems, but would not reduce the metallicity of low $\NHI$ systems
because it would make the star formation even more efficient than 
it is in current simulations.  If future cosmological simulations 
with a more sophisticated modelling of star formation and 
SN feedback confirm the existence of high-metallicity, high-$\NHI$ 
systems, then they could turn into an interesting challenge for 
the CDM model.

Another important common feature found in both simulation studies is
the mild evolution of the mean metallicity of DLAs with redshift. It
is encouraging that the two entirely different simulation
methodologies agree on the rate of this evolution, which in turn is
consistent with the observed rate.  Note that processes such as galaxy
mergers, gas infall, and outflows all play an important role in the
chemical evolution of DLAs, and these processes, which often cannot
be treated accurately in semi-analytic models of the chemical
evolution of DLAs, are included dynamically and self-consistently in
both simulations.

In conclusion, together with our previous results presented in
\citet{Nag03}, we have shown that the DLAs found in our simulation
series have many plausible properties. In particular, they are in good
agreement with recent observations of the total neutral \HI mass
density, the \HI column density distribution function, the abundance
of DLAs, and the distribution of SFR in DLAs.  However, our simulated
DLAs show typically considerably higher metallicity than what is
presently observed for the bulk of these systems. This likely
indicates that metal transport and mixing processes have not been
efficient enough in our simulations. It will be interesting to
study more sophisticated metal enrichment models in future simulations
in order to further improve our understanding of the nature of DLAs in
hierarchical CDM models.


\section*{Acknowledgments}
K.N. thanks for the hospitality of the {\it Max-Planck-Institut f\"{u}r 
Astrophysik} in Garching where part of this work was completed.
We thank Art Wolfe and Jason Prochaska for stimulating discussions, 
Simon White for the comments on the star-to-gas mass ratio, 
Jason Prochaska and Max Pettini for providing us with their data points 
in Figure~\ref{metal_bulk.eps}.
This work was supported in part by NSF grants ACI96-19019, AST 98-02568, 
AST 99-00877, and AST 00-71019 and NASA ATP grant NAG5-12140.  
The simulations were performed at the Center for Parallel Astrophysical 
Computing at the Harvard-Smithsonian Center for Astrophysics.


\bsp

\label{lastpage}


\begin{thebibliography}{}

\bibitem[\protect\citeauthoryear{Adelberger et al.}{1998}]{Ade98} 
	Adelberger K. L., Steidel C. C., Giavalisco M., Dickinson M., 
	Pettini M., Kellogg M., 1998, ApJ, 505, 18

\bibitem[\protect\citeauthoryear{Bahcall}{1984}]{Bah84}
	Bahcall J. N., 1984, ApJ, 287, 926 

\bibitem[\protect\citeauthoryear{Bahcall, Flynn, \& Gould}{1992}]{Bah92}
	Bahcall J. N., Flynn C., Gould A., 1992, ApJ, 389, 234

\bibitem[\protect\citeauthoryear{Baugh et al.}{1998}]{Bau98} Baugh C. M.,
	Cole S., Frenk C. S., Lacey C. G., 1998, ApJ, 498, 504 

\bibitem[\protect\citeauthoryear{Becker et al.}{2001}]{Becker01} 
        Becker R. H., et al. 2001, AJ, 122, 2850

\bibitem[\protect\citeauthoryear{Binney \& Merrifield}{1998}]{Binney98} 
	Binney J., Merrifield M., 1998, {\it Galactic Astronomy}, 
	Princeton University Press, p.656-664

\bibitem[\protect\citeauthoryear{Blain et al.}{1999}]{Bla99} Blain A. W.,
	Smail I., Ivison R. J., Kneib J.-P., 1999, MNRAS, 302, 632 

\bibitem[\protect\citeauthoryear{Blanton et al.}{1999}]{Blanton99} 
	Blanton M., Cen R., Ostriker J. P., Strauss M. A., 
	1999, ApJ, 522, 590

\bibitem[\protect\citeauthoryear{Boiss\'e et al.}{1998}]{Boisse98} 
	Boiss\'e P., Le Brun V., Bergeron J., Deharveng J. M., 
	1998, A\&A, 333, 841

\bibitem[\protect\citeauthoryear{Bromm et al.}{2003}]{Bromm03}
	Bromm V., Yoshida N., Hernquist L., 2003, ApJ, submitted. (astro-ph/0305333)

\bibitem[\protect\citeauthoryear{Cen et al.}{2002}]{Cen02} Cen R., Ostriker J. P., 
	Prochaska J. X., Wolfe A. M., 2002, preprint (astro-ph/0203524)

\bibitem[\protect\citeauthoryear{Cr\'{e}z\'{e} et al.}{1998}]{Cre98} Cr\'{e}z\'{e} M., Chereul E., Bienaym\'{e} O., Pichon C., 1998, A\&A, 329, 920

\bibitem[\protect\citeauthoryear{Croft et al.}{2001}]{Croft01} Croft R. A. C.,
        Di Matteo T., Dav\'e R., Hernquist L., Katz N., Fardal M. A.,
        Weinberg D.H., 2001, ApJ, 557, 67

\bibitem[\protect\citeauthoryear{Dame}{1993}]{Dame} Dame T. M., 1993, 
	{\it Back to the Galaxy}, AIP Conference Proceedings 278, eds. 
	Holt, S. S., Verter, F. (New York: AIP), p.267 

\bibitem[\protect\citeauthoryear{Dav\'e et al.}{1999}]{Dave99} Dav\'e R., 
	Hernquist L., Katz N., Weinberg D. H., 1999, ApJ, 511, 521

\bibitem[\protect\citeauthoryear{Dijkstra et al.}{2003}]{Dij} Dijkstra M., 	
	Haiman Z., Rees M. J., \& Weinberg D. H. 2003 (astro-ph/0308042)

\bibitem[\protect\citeauthoryear{Ellison et al.}{2001}]{Ellison} 
	Ellison S. L., Yan L., Hook I. M., Pettini M., Wall J. V., Shaver P.
	2001, A\&A, 379, 393

\bibitem[\protect\citeauthoryear{Flynn \& Fuchs}{1994}]{Flynn94}
	Flynn C., Fuchs B., 1994, MNRAS, 270, 471 

\bibitem[\protect\citeauthoryear{Gardner et al.}{1997a}]{Gar97a} Gardner J. P., 
	Katz N., Hernquist L., Weinberg D. H., 1997a, ApJ, 484, 31

\bibitem[\protect\citeauthoryear{Gardner et al.}{1997b}]{Gar97b} Gardner J. P., 
	Katz N., Weinberg D. H., Hernquist L., 1997b, ApJ, 486, 42

\bibitem[\protect\citeauthoryear{Gardner et al.}{2001}]{Gar01} Gardner J. P., 
	Katz N., Hernquist L., Weinberg D. H., 2001, ApJ, 559, 131

\bibitem[\protect\citeauthoryear{Haardt \& Madau}{1996}]{Haa96} Haardt F., 
	Madau P., 1996, ApJ, 461, 20 

\bibitem[\protect\citeauthoryear{Hernquist}{1993}]{Her93} 
	Hernquist L., 1993, ApJ, 404, 717

\bibitem[\protect\citeauthoryear{Hernquist \& Springel}{2003}]{Her03} 
	Hernquist L., Springel V., 2003, MNRAS, 341, 1253

\bibitem[\protect\citeauthoryear{Holmberg \& Flynn}{2000}]{Hol00}
	Holmberg J., Flynn C., 2000, MNRAS, 313, 209

\bibitem[\protect\citeauthoryear{Jimenez et al.}{1999}]{Jim99} 
	Jimenez R., Bowen D. V., Matteucci F., 1999, ApJ, 514, L83

\bibitem[\protect\citeauthoryear{Katz, Hernquist \& Weinberg}{1999}]{KHW99} 
        Katz N., Hernquist L., Weinberg D. H., 1999, ApJ, 523, 463

\bibitem[\protect\citeauthoryear{Katz et al.}{1996}]{Katz96} Katz N., 
	Weinberg D. H., Hernquist L., 1996, ApJS, 105, 19 

\bibitem[\protect\citeauthoryear{Kauffmann}{1996}]{Kau96} Kauffmann G.,
	1996, MNRAS, 281, 475

\bibitem[\protect\citeauthoryear{Kauffmann et al.}{1999}]{Kau99} 
        Kauffmann G. A. M., Colberg J. M., Diaferio A., White S. D. M., 1999,
        MNRAS, 307, 529

\bibitem[\protect\citeauthoryear{Kennicutt}{1998}]{Ken98} 
	Kennicutt, R. C. Jr., 1998, ARA\&A, 36, 189

\bibitem[\protect\citeauthoryear{Kravtsov}{2003}]{Kra03} Kravtsov A. V.
	2003, preprint (astro-ph/0303240) 

\bibitem[\protect\citeauthoryear{Kuijken \& Gilmore}{1991}]{Kui91} 
	Kuijken K., Gilmore G., 1991, ApJ, 367, L9

\bibitem[\protect\citeauthoryear{Kuijken \& Gilmore}{1989}]{Kui89} 
	Kuijken K., Gilmore G., 1989, MNRAS, 239, 605 

\bibitem[\protect\citeauthoryear{Laird et al.}{1988}]{Laird98} 
	Laird J. B., Rupen M. P., Carney B. W., Latham D. W. 1988, AJ, 96, 1908

\bibitem[\protect\citeauthoryear{Lanfranchi \& Friaca}{2003}]{Lanf03} 
	Lanfranchi G. A., Friaca A. C. S., 2003, MNRAS, in press (astro-ph/0304119)

\bibitem[\protect\citeauthoryear{Lanzetta et al.}{2002}]{Lan02} Lanzetta K. M.,
	Yahata N., Pascarelle S., Chen H-W., Fern{\'a}ndez-Soto A., 2002, 
	ApJ, 570, 492 

\bibitem[\protect\citeauthoryear{Lanzetta}{1993}]{Lan93}
        Lanzetta K. M. 1993, in The Environment and Evolution of Galaxies,
	eds. J. M. Shull \& H. A. Thronson (Dordrecht: Kluwer), p237

\bibitem[\protect\citeauthoryear{Lu et al.}{1996}]{Lu96} Lu L, Sargent W. L. W.,
	Barlow T. A., Churchill C. W., Vogt S., 1996, ApJS, 107, 475

\bibitem[\protect\citeauthoryear{Lucy}{1977}]{Lucy77} 
	Lucy L., 1977, AJ, 82, 1013

\bibitem[\protect\citeauthoryear{Mac Low \& Ferrara}{1999}]{MacLow}
	Mac Low, M. M. \& Ferrara, A., 1999, ApJ, 513, 142

\bibitem[\protect\citeauthoryear{Maller, Prochaska, Somerville, \& Primack}{2002}]{Maller02} Maller A. H., Prochaska J. X., Somerville R. S., Primack J. R., 2002, MNRAS, submitted (astro-ph/0211231)

\bibitem[\protect\citeauthoryear{Maller, Prochaska, Somerville, \& Primack}{2001}]{Maller01} Maller A. H., Prochaska J. X., Somerville R. S., Primack J. R., 2001, MNRAS, 326, 1475

\bibitem[\protect\citeauthoryear{McKee \& Ostriker}{1977}]{McKee77}
	McKee C. F., Ostriker, J. P., 1977, ApJ, 218, 148 

\bibitem[\protect\citeauthoryear{Mo, Mao, \& White}{1999}]{Mo99} 
	Mo H. J., Mao S., White S. D. M., 1999, MNRAS, 304, 175

\bibitem[\protect\citeauthoryear{Mo \& Fukugita}{1996}]{Mo96} Mo H. J., 
	Fukugita, 1996, ApJ, 467, L9

\bibitem[\protect\citeauthoryear{Nagamine, Springel, \& Hernquist}{2003}]{Nag03} 
	Nagamine K., Springel V., Hernquist L., 2003, MNRAS, submitted
	(astro-ph/0302187)

\bibitem[\protect\citeauthoryear{Nagamine}{2002}]{Nag02} Nagamine K., 
	2002, ApJ, 564, 73

\bibitem[\protect\citeauthoryear{Nagamine et al.}{2001}]{Nag01a} Nagamine K.,
	Fukugita M., Cen R., Ostriker J. P., 2001, ApJ, 558, 497 

\bibitem[\protect\citeauthoryear{Pascarelle, Lanzetta, \& Fern{\'a}ndez-Soto}{1998}]{Pas98} Pascarelle S., Lanzetta K. M., Fern{\'a}ndez-Soto A., 1998, ApJ, 508, L1 

\bibitem[\protect\citeauthoryear{Pearce et al.}{1999}]{Pearce99} 
	Pearce et al., 1999, ApJ, 521, 99

\bibitem[\protect\citeauthoryear{Pei \& Fall}{1995}]{Pei95} 
	Pei Y. C. \& Fall S. M. 1995, ApJ, 454, 69 
							  
\bibitem[\protect\citeauthoryear{Pettini}{2003}]{Pettini03} Pettini M. 
        2003, XIII Canary Islands Winter School of Astrophysics,
	{\it `Cosmochemistry: The Melting Pot of Elements'} 
        (astro-ph/0303272)

\bibitem[\protect\citeauthoryear{Pettini et al.}{1999}]{Pettini99}
        Pettini M., Ellison S. L., Steidel C. C., Bowen D. V.
        1999, ApJ, 510, 576 

\bibitem[\protect\citeauthoryear{Pettini et al.}{1994}]{Pettini94}
	Pettini M., Smith L. J., Hunstead, R. W., King D. L., 
	1994, ApJ, 426, 79

\bibitem[\protect\citeauthoryear{Prantzos \& Boissier}{2000}]{Pra00}
	Prantzos N., Boissier S., 2000, MNRAS, 315, 82

\bibitem[\protect\citeauthoryear{Prochaska et al.}{2003}]{Pro03} 
	Prochaska J. X., Gawiser E., Wolfe A. M., Castro S., Djorgovski G., 
	2003, ApJ, submitted (astro-ph/0305314)

\bibitem[\protect\citeauthoryear{Prochaska \& Wolfe}{2002}]{Pro02} 
	Prochaska J. X. \& Wolfe A. M., 2002, ApJ, 566, 68

\bibitem[\protect\citeauthoryear{Prochaska \& Wolfe}{2000}]{Pro00} 
	Prochaska J. X. \& Wolfe A. M., 2000, ApJ, 533, L5

\bibitem[\protect\citeauthoryear{Prochaska \& Wolfe}{1999}]{Pro99} 
	Prochaska J. X. \& Wolfe A. M., 1999, ApJS, 121, 369

\bibitem[\protect\citeauthoryear{Prochaska \& Wolfe}{1998}]{Pro98} 
	Prochaska J. X. \& Wolfe A. M., 1998, ApJ, 507, 113
	
\bibitem[\protect\citeauthoryear{Schaye}{2001}]{Schaye} 
	Schaye, J. 2001, ApJ, 562, L95

\bibitem[\protect\citeauthoryear{Shapley et al.}{2001}]{Sha01} 
	Shapley A. E., Steidel C. C., Adelberger K. L., Dickinson M., 
	Giavalisco M., Pettini M., 2001, ApJ, 562, 95 

\bibitem[\protect\citeauthoryear{Sokasian et al.}{2003}]{Sok03}
        Sokasian A., Abel T., Hernquist L., Springel V., 2003, MNRAS, in press
        (astro-ph/0303098)

\bibitem[\protect\citeauthoryear{Somerville, Primack, \& Faber}{2001}]{Som01}
	Somerville R. S., Primack J. R., Faber S. M., 2001, MNRAS, 320, 504  

\bibitem[\protect\citeauthoryear{Springel \& Hernquist}{2002}]{SH02} 
	Springel V., Hernquist L., 2002, MNRAS, 333, 649

\bibitem[\protect\citeauthoryear{Springel \& Hernquist}{2003a}]{SH03a} 
	Springel V., Hernquist L., 2003a, MNRAS, 339, 289

\bibitem[\protect\citeauthoryear{Springel \& Hernquist}{2003b}]{SH03b} 
	Springel V., Hernquist L., 2003b, MNRAS, 339, 312

\bibitem[\protect\citeauthoryear{Springel, Yoshida \& White}{2001}]{Gadget} 
        Springel V., Yoshida N., White S.~D.~M., 2001, New Astronomy, 6, 79

\bibitem[\protect\citeauthoryear{Steidel et al.}{1999}]{Ste99} Steidel C. C., 
	Adelberger K. L., Giavalisco M., Dickinson M., Pettini M., 1999, 
	ApJ, 519, 1

\bibitem[\protect\citeauthoryear{Storrie-Lombardi \& Wolfe}{2000}]{Sto00} 
	Storrie-Lombardi L. J., Wolfe A. M.,  2000, ApJ, 543, 552

\bibitem[\protect\citeauthoryear{Theuns et al.}{2002}]{Theuns02} Theuns T.,
	Viel M., Kay S., Schaye J., Carswell R. F., Tzanavaris P. 2002,
	ApJ, 578, L5

\bibitem[\protect\citeauthoryear{Wolfe, Prochaska, \& Gawiser}{2003a}]{Wol03a} 
	Wolfe A. M., Prochaska J. X., Gawiser E., 2003a, ApJ, in press 
	(astro-ph/0304040)

\bibitem[\protect\citeauthoryear{Wolfe, Prochaska, \& Gawiser}{2003b}]{Wol03b} 
	Wolfe A. M., Prochaska J. X., Gawiser E., 2003b, preprint (astro-ph/0304042)

\bibitem[\protect\citeauthoryear{Wolfe et al.}{1995}]{Wol95} Wolfe A. M., 
	Lanzetta K. M., Foltz C. B., 1995, ApJ, 454, 698

\bibitem[\protect\citeauthoryear{Wolfe et al.}{1986}]{Wol86} Wolfe A. M., 
	Turnshek D. A., Smith H. E., Cohen R. S., 1986, ApJS, 61, 249 

\bibitem[\protect\citeauthoryear{Weinberg, Hernquist, \& Katz}{2002}]{Wei02} 
	Weinberg D. H., Hernquist L., Katz N., 2002, ApJ, 571, 15

\bibitem[\protect\citeauthoryear{Yoshida et al.}{2002}]{Yoshida02} Yoshida N, 
	Stoehr F., Springel V., White S. D. M., 2002, MNRAS, 335, 762

\bibitem[\protect\citeauthoryear{Wyse \& Gilmore}{1995}]{Wyse95}
	Wyse R. F. G. \& Gilmore G. 1995, AJ, 110, 2771 

\end{thebibliography}
\end{document}